%% file: main.tex
\documentclass{aa}

\usepackage[english]{babel}
\usepackage{amssymb}
\usepackage{mathpazo}
\usepackage{graphicx}
\usepackage{epstopdf}
\usepackage{epsfig}
\usepackage{natbib}
\usepackage{hyperref}
\usepackage{multirow}
\usepackage{textcomp}
\usepackage{lscape}
\usepackage{subfigure}
\usepackage{rotating}
\usepackage{pdfpages}
\usepackage{amsmath}
\usepackage{amssymb,color,latexsym,longtable,array}
\usepackage{txfonts} 
\usepackage{lscape}
\usepackage{afterpage}
\bibpunct{(}{)}{;}{a}{}{,}
\graphicspath{{Images/}}
\usepackage[labelfont=bf, textfont=sl]{caption}

\newcommand*{\rom}[1]{\uppercase\expandafter{\romannumeral #1\relax}}
\usepackage{setspace}

\begin{document}

\newif\ifreftype
\reftypefalse
\title{Determining Spectroscopic Redshifts by Using k Nearest Neighbors
Regression}
\subtitle{\rom{1}. Description of Method and Analysis}
\author{
S. D. K\"ugler\inst{1} \and
K. Polsterer\inst{1} \and
M. Hoecker\inst{1}
}
\institute{
Heidelberger Institut f\"ur Theoretische Studien (HITS),
Schloss-Wolfsbrunnenweg 35, D-69118 Heidelberg, Germany\\
email: dennis.kuegler@h-its.org
} 

\date{Received/Accepted}

\abstract{In astronomy, new approaches to process and analyze the
  exponentially increasing amount of data are inevitable. For spectra,
  such as in the Sloan Digital Sky Survey spectral database, usually
  templates of well-known classes are used for classification. In case 
  the fitting of a template fails, wrong spectral properties (e.g. redshift)
  are derived. Validation of the derived properties is the key to
  understand the caveats of the template-based method. }
  {In this paper we present a method to statistically compute the redshift
  $z$ based on a similarity approach. This allows us to determine redshifts in
  spectra for emission and absorption features without using any
  predefined model. Additionally we show how to determine the redshift
  based on single features. As a consequence we are, e.g. able to filter
  objects which show multiple redshift components.} 
  {The redshift calculation is performed by
  comparing predefined regions in the spectra and applying a nearest
  neighbor regression model for every predefined emission and absorption region,
  individually.} {The choice of the model parameters controls the quality and
  the completeness of the redshifts. For $\approx$90\%
  of the analyzed 16,000 spectra of our reference and test sample a certain
  redshift can be computed which is comparable to the completeness of SDSS
  ($96$\%).
  The redshift calculation yields a precision for every individually tested
  feature that is comparable with the overall precision of the redshifts of
  SDSS. Using the new method to compute redshifts we could additionally
  identify 14 spectra with a significant shift between emission and absorption
  or emission and emission lines.
  The results show already the immense power of this simple machine learning
  approach for investigating huge databases such as the SDSS.}{}

\keywords{Galaxies: distances and redshifts --
Astronomical data bases --
Catalogs --
Methods: data analysis --
Methods: statistical}

\maketitle

\input{Introduction.tex}
\input{Data.tex}
\input{Analysis.tex}
\input{Results.tex}

\input{Discussion.tex}

\begin{acknowledgements}
\thanks{Funding for SDSS-III has been provided by the Alfred P. Sloan
Foundation, the Participating Institutions, the National Science
Foundation, and the U.S. Department of Energy Office of Science.
The SDSS-III web site is http://www.sdss3.org/.
SDSS-III is managed by the Astrophysical Research Consortium for the
Participating Institutions of the SDSS-III Collaboration including the
University of Arizona, the Brazilian Participation Group, Brookhaven
National Laboratory, Carnegie Mellon University, University of Florida,
the French Participation Group, the German Participation Group, Harvard
University, the Instituto de Astrofisica de Canarias, the Michigan
State/Notre Dame/JINA Participation Group, Johns Hopkins University,
Lawrence Berkeley National Laboratory, Max Planck Institute for Astrophysics,
Max Planck Institute for Extraterrestrial Physics, New Mexico State University,
New York University, Ohio State University, Pennsylvania State University,
University of Portsmouth, Princeton University, the Spanish Participation Group,
University of Tokyo, University of Utah, Vanderbilt University, University of
Virginia, University of Washington, and Yale University.}

\thanks{The authors thank Michael Schick for very fruitful discussions on the
topic of uncertainty quantification.}
\thanks{SDK would like to thank the Klaus Tschira Foundation for their financial
support.}
\end{acknowledgements}
\bibliographystyle{aa}
\bibliography{library.bib}
\clearpage
\onecolumn
\input{Appendix.tex}
\end{document}

%% file: Introduction.tex
\section{Introduction}
In the past decades the rapidly increasing amount of available data has been
one of the greatest challenges in astronomy. In contrast to the amount of data,
the number of techniques and the knowledge how to analyze these large data sets 
increased only slowly over time. When the first digital, photometric all-sky 
surveys were performed, the amount of available data was already too large
to be inspected manually. With the advent of spectroscopic surveys and
additional photometric surveys in multiple wavelengths, the available data
volume increased so rapidly that novel approaches are mandatory.


So far the most successful survey in astronomy is the Sloan Digital Sky Survey
(SDSS, \citealp{2000AJ....120.1579Y}) which contains in its current 10th data
release (DR10, \citealp{2014ApJS..211...17A}) photometry for one billion objects
and spectra covering the near-UV to the near-IR for roughly three million
objects. 
In future, surveys such as the Large Sky Area Multi-Object Fiber
Spectroscopic Telescope (LAMOST, \citealp{2012RAA....12.1197C})
will reach this amount of data in a fraction of the time needed by SDSS.
Thus more advanced techniques for handling those immense data streams have to
be developed.

The determination of spectral redshifts and classifications of the SDSS spectra
is based on template fitting. Therefore generalized templates are created
by combining spectra of similar objects for all empirically determined classes
of objects. By fitting those templates to the spectra, a number of predefined
properties, e.g. redshift, can be individually computed for every object. By
applying all available templates to the data while allowing for some variation
in a set of parameters, e.g. width of features, and testing the reliability of
every model by computing a reduced $\chi^2$, the best fitting template is
determined. 
Instead of using the full information available, just a simplified model with a
limited flexibility is applied which does not allow a more detailed
discussion of individual properties.
Furthermore the choice of the reference spectra and the creation of these
templates has a strong impact on the determined properties. 

With this publication we want to emphasize the power of statistical learning
in huge spectral databases. Hereby, huge refers to a large number of entities
and dimensions. While this approach can principally be applied to any database,
we focus on SDSS.
There are many applications of machine learning techniques in astronomy
\citep[see][]{2009arXiv0911.0505B,2010IJMPD..19.1049B}. So far spectroscopically
derived properties have been mainly used as ground truth to, e.g. estimate
redshifts on photometric data
\citep[see][]{2011MNRAS.418.2165L, 2011arXiv1108.4696G, 2013MNRAS.428..226P}.
In contrast less attention has been paid to the application of machine
learning to the spectral data itself 
\citep[see][]{2009ApJ...691...32R,2012A&A...541A..77M} which can be mainly
attributed to the ``curse of dimensionality'' \citep[see][]{Bellman1961}.
The ultimate goal would be to obtain spectral properties which are not based on
the created templates but on the rich experience existing in the database
instead.

The algorithm presented in this paper will perform a consistency check
of the redshift calculated by the SDSS pipeline. We therefore assume that the
majority of the spectra is fairly well described by one of the templates and
thus the redshift is determined reasonably precise. Of course the templates
do not describe all kinds of objects perfectly, thus at least some will be
misfit. The great improvement in calculating redshifts based on a data-driven
approach is that the redshifts can be determined model-independent. This method
is suitable for determining redshifts of unknown spectra and in a forth-coming
paper we will present a value-added catalog of redshifts to the existing SDSS
spectra. In this paper we will focus on the technical side and explain the
impact of the choice of different model parameters. In order to highlight the
power of this new method, some outliers in terms of redshift in the used
subsample are presented.
The motivation for employing new methods for redshift computation is manifold:
\begin{enumerate}
  \item \emph{Validation}:
{Cross-validating the self-consistency of the computed redshifts is
crucial to understand caveats of the SDSS pipeline. The independent
determination of a redshift increases the confidence and the number of
reliable redshifts. }\newline
  \item \emph{Calculating redshifts}:
{We are able to determine model-independent redshifts of existing and future
spectra with high precision.
This is possible since we are determining the redshift as an ensemble property
and thus the theoretical resolution can be improved statistically with
the number of similar spectra in the reference database as well as with the
dimension of the feature vector.}\newline
  \item \emph{Rare objects}:
{Many different attempts have been performed to find rare objects in the
SDSS spectral database which show shifts between spectral features
\citep[see][]{2004AJ....127.1860B,2011ApJ...738...20T}. With the presented
method we will be able to detect more of those since our method can deal with
lower S/N than the template fits.}\newline
  \item \emph{Unexpected behavior}:
{This can be caused by objects of a previously unknown class or by a
superposition of two classes. Those objects might possibly be the science
drivers in the near future. Also artifacts in the reduction pipeline/in
the data can be discovered.}\newline

\end{enumerate} 


The paper is structured as follows: \S2 describes the data used for creating and
testing our model. In \S3 we will explain the basic approach used in
our method in more detail. In \S4 we discuss the performance of our method in
terms of precision and reliability. Also some outliers and peculiar objects are
discussed in more detail. A summary and an outlook will follow in \S5. In a
follow-up paper we describe the value-added catalog which gives redshifts for
all available objects based on specific spectral regions.
Additionally a catalog containing all detected outliers will be presented there.

%% file: Data.tex
\section{The SDSS Spectroscopic Database}
For testing our method we are analyzing the spectroscopic database of SDSS. This
survey uses a dedicated 2.5m mirror telescope located at the Apache Point
Observatory (New Mexico, USA) to map the northern galactic cap and is a joint
project by USA, Japan, Korea and Germany.

The telescope was first used to image different stripes of the 
northern hemisphere in 5 filter bands using the drift scan method. 
Subsequently interesting objects were selected 
by brightness limits and different colors cuts for spectroscopy
 (\mbox{$R = \lambda/\Delta \lambda \approx 2,000$}) with $3,600\,\AA\leq
 \lambda \leq10,000\,\AA$ \citep{2001AJ....122.2267E, 2002AJ....123.2945R, 2002AJ....124.1810S}.
 Note that those selection criteria directly have an impact on the quality of
 reference sample. 
%
In the current DR10 \citep{2014ApJS..211...17A} more than 3
million spectra were taken of which far more than 2 million are non-stellar
sources according to the SDSS-classification.

It is important to mention that depending on the applied learning technique a
large number of reference objects with a representative sampling is mandatory.
With millions of objects, the SDSS is more than sufficiently
large.\footnote{This statement is not only valid for the in-sample method
presented here but also for the application on other datasets, {\bf as long as
the wavelength coverage and the target selection criteria are comparable}. This
is because the data complexity of the reference sample, SDSS in this case, remains
the same and thus a comparable number of references is needed for a similiar
precision.}

\subsection{Data Calibration/SDSS pipeline}
As mentioned in the caveats of SDSS, the night sky subtraction
can suffer from severe inaccuracy by rapidly changing conditions, e.g. auroral
activity. Thus the night sky subtraction leaves a severe signature in some of
the spectra, which is sometimes not taken into account correctly in the error
estimation. As a consequence faint features in the vicinity of strong night sky
emission lines might be artifacts.
The spectra are automatically labeled, flux as well as wavelength calibrated,
and eventually combined with potentially pre-existing observed spectra of the
same object.

In a second step the calibrated spectra were processed via an  
identification pipeline which assigned a redshift, a classification 
and a velocity dispersion to the individual spectra \citep{2012AJ....144..144B}.
The classification and redshift determination was hereby performed 
with a principal component analysis (PCA) of a rest-frame shifted 
training sample. A linear combination of eigenspectra were then shifted 
with respect to flux and wavelength until a minimal residual was 
reached. 
The precision of the redshift for a single line
is limited by the resolution per pixel (\mbox{$\sim100\,\text{kms}^{-1}$}) of
the spectrograph but can be improved by computing it independently for all lines
available. This method is extremely efficient for spectra which show the
expected behavior and as confirmed by performing a self-consistency check later
on, the quality of the SDSS redshifts has a high reliability.

\subsection{Reference And Test Sample}
\begin{figure}
\centering
\caption{Comparison of the redshift distribution of the selected
subsample (green) and the entire SDSS (blue). Note a steep drop in the frequency
towards redshifts $z>0.25$. Single redshift bins are apparently
undersampled.
\label{fig:redDist}}
\includegraphics[width=0.5\textwidth]{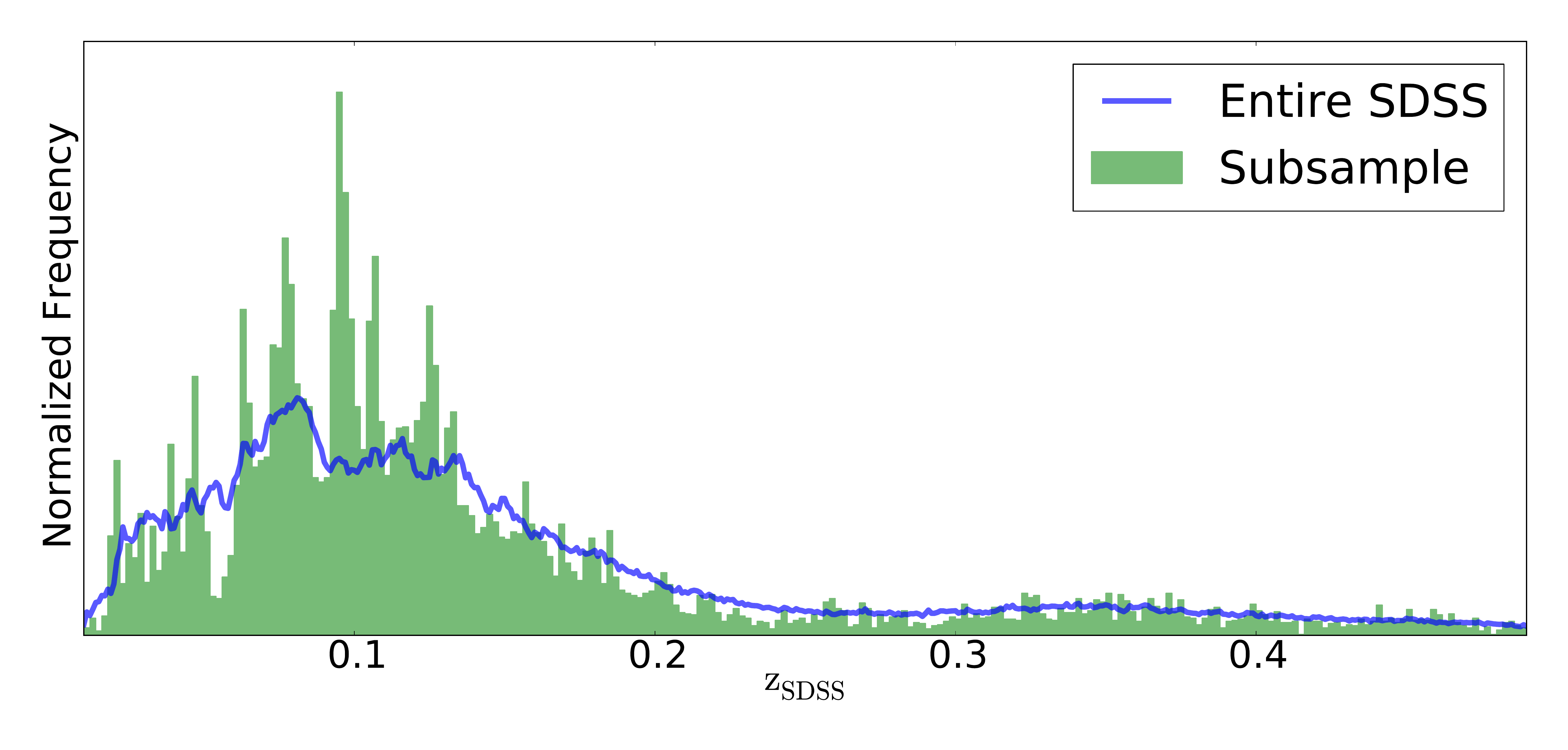}
\end{figure}

The analysis of the method was performed on a small subsample of the
SDSS data in order to make the different model and parameter evaluations
computationally feasible. The analysis of the algorithm is limited to the plates
0266 to 0289 including the exposures of all modified Julian dates (MJDs).
Additionally the sample was restricted to the redshift range between $0.01\leq z
\leq 0.5$. The selected restriction allows a more reliable prediction of the
regression value, as the density of reference targets in the direct neighborhood
is sufficiently high.
The chosen subsample includes 16,049 spectra in total. The redshift
distribution of the spectra can be found in \mbox{Figure \ref{fig:redDist}}.
In the following, this sample will be used as reference and test set at the
same time, i.e. we will perform a leave-one-out cross validation. That means
that all but the target spectrum are reference spectra. As we are only able
to compute redshifts within the covered feature space, under-represented
objects (high-redshifted galaxies, QSO) will yield worse redshifts than
normally represented redshifts.

%% file: Analysis.tex
\section{Applied Method}
The basic idea for determining the spectroscopic redshift $z$ is to perform a
comparison between similar objects. This is done by
finding objects which look similar in terms of Euclidean distance and then
computing the regression value of the unknown target by comparing it to the
redshifts of the most similar spectra.

To be able to compare the spectra, instead of using the plain SDSS spectra we
have to pre-process them.

The method is a purely data-driven approach without deriving a
generalization and thus the quality of the redshifts relies directly on the
chosen reference sample. While this seems contradictory on first sight the
method performs comparably on a smaller but representative reference set. It is
obvious that the choice of a representative reference sample can just be
obtained when domain-knowledge is included. Limiting the reference sample in
redshift space would limit the derived values, respectively.


\subsection{k Nearest Neighbors (kNN) Regression}
Our method is based on $k$ nearest neighbor ($k$NN) regression which is
a commonly used technique in statistical learning \citep{Hastie2009}.
All spectra have $d$ datapoints (corresponding to the individual
flux measurements in the spectra) and are thus members of a
\mbox{$d$-dimensional} feature space.
The reference sample $\mathcal{R}$ consists of $m$ entities which
corresponds to the number of reference spectra from which the
model learns, 16,048 in this case.
Mathematically this sample can be described with,
\begin{equation}
\mathcal{R} =
\left(\left(\overline{x}_1,y_1\right),\,\ldots,\,
\left(\overline{x}_{m},{y_{m}}\right)\right)
\,\in \mathbb{R}^{d} \times \mathbb{R}
\end{equation}
\noindent where $\overline{x}_i$ is the $i$-th $d$-dimensional input vector
(spectrum under consideration) corresponding to the flux value in each
pixel and $y_i$ is the redshift value $z$ assigned by the SDSS pipeline.

The $k$NN regression is based on calculating similarities in the $d$-dimensional
feature space. For any $d$-dimensional feature vector $\overline{s}$
the similarity to a reference object $\overline{x}_i$ can be estimated with 
distance measure $\Delta\left(\overline{x}_i,\overline{s}\right)$.
The most commonly used metrics are:
\begin{equation*}
\Delta\left(\overline{x}_i,\overline{s}\right) =
\left(\sum_{j=1}^d~\lvert{{\overline{x}_i}_j-\overline{s}_j}\rvert^p\right)^{1/p}
:=
\begin{cases} 
\text{Manhattan} & \text{for } p=1\\
\text{Euclidean} & \text{for } p=2\\
\text{Minkowski} & \text{otherwise}
\end{cases}
\end{equation*}
\newline
\noindent The impact of the choice of the metric on the final results was
only marginal. Therefore we solely use the common Euclidean distance. 
In general the neighbourhood $N_k(\overline{s})$ is determined on the basis of
the representation of the reference objects $\overline{x}_i$ in the
feature space, such that

\begin{equation}
y(\overline{s})=\frac{1}{k}\sum \limits_{\overline{x}_i \epsilon
 N_k(\overline{s})} y_i = \underset{{\overline{x}_i \epsilon
 N_k(\overline{s})}}{\mathrm{mean}}\left(y_i\right)
 \end{equation}

\noindent however here we make use of a modified version:

\begin{equation}
y(\overline{s})=\underset{{\overline{x}_i \epsilon
 N_k(\overline{s})}}{\mathrm{median}}\left(y_i\right)
 \label{eq:median}
\end{equation}

\noindent 
For finding the $k$ most similar spectra $N_k(\overline{s})$ different
algorithms exist. The most straight-forward one being the brute-force method
where simply every spectrum is compared to each other and the distance is
computed. In contrast, spatial structures exist (kd-, ball-trees) which
are able to structure the data in advance. Thereby the average time to find
the closest spectra is significantly lower once the search structure is created.
When experimenting with spatial trees we learned that apparently the
dimension of our data is so high and the data themselves are so unstructured
that spatial trees do not perform significantly better than the brute-force
method and as a consequence only the brute force method is used throughout the
paper and for the future catalog.


The considered $k$NN regression is limited to interpolation of values within the
reference sample. As a consequence redshifts of objects with extremely high
redshift or very peculiar spectral features can not be determined correctly.

\subsection{Requirements}
\label{subsect:prereq}
The method of $k$NN regression can only work efficiently if the following
requirements are met:
\begin{enumerate}
  \item The majority of redshift determinations by SDSS is
  correct:\newline\newline {In the following the deviation of the SDSS redshifts
 in comparison to the correct redshift is assumed to be small. This is
verified by comparing our results to the redshifts determined by SDSS. One
has to keep in mind that for a large fraction of the data the template fitting
works quite well and the redshifts are fairly reliable.}\newline
  \item The number of objects in the reference data set is large compared to the
  dimensionality:\newline\newline 
{This is already met in our test sub-sample. Nonetheless this is quite
surprising as the number of entities is in the order of the number of dimensions
(4,000). It appears that the multi-dimensional feature space is sufficiently
homogeneously populated with reference objects. Applying this method to the
entire database will just strengthen that assumption further.}\newline
  \item It is possible to distinguish noise from real signals for most of the
  data:\newline\newline 
  {This requirement is harder to meet as the
  distinction between signals and noise, especially for low S/N
  spectral lines, has always been a huge challenge for astronomers.
  In this work we will use an approach that is based on a simple similarity
  measure used by the type of the applied regression method. The basic
  assumption is: When a detectable line exists anywhere in the spectrum it
  should be possible to find similar spectra which, within their errors, have a
  similar redshift. Those form a sharp distribution around the real value. On
  the other hand a  spectrum that contains pure noise will yield an even
  distribution of redshift values over the entire tested redshift range and
  thus the average deviation from the median/mean will be quite high.
  In the distribution of so called errors, which correspond to the deviation of
  reference redshifts across similar spectra, one would naively expect a
  superposition of two behaviors:
  The dominant component is a distribution which shows a drop towards higher
  deviations with a width that is comparable to the sensitivity of the method.
  This distribution corresponds to redshifts based on true absorption/emission
  lines. Underlying to the first component there is a flatter distribution
  representing the spectra which contain mostly noise.
  This will be further discussed in \mbox{Section \ref{experiments}}.}
\end{enumerate}

\subsection{Pre-processing}

The pre-processing is needed to make the spectra comparable. Effects like
apparent brightness are not important, since we are solely interested in
absorption and emission features. Therefore the behavior of the continuum has
to be estimated and subtracted.

\subsubsection{Regridding}
The dispersion resolution between different fibers on a single plate and between
the plates themselves differ slightly. In order to always be able to compare the
correct wavelength bins, which do not exactly agree with redshift bins, the
spectra have to be regridded. We are therefore creating a global grid which
is defined by:
\begin{equation}
\log\left(\lambda\left(p\right)\right) = 0.0001 \cdot p+3.5222
\end{equation}
where $\lambda$ is the wavelength in $\AA$ for a given pixel position $p$, with
$0\leq p < 5,100$. The parameters of the function are chosen such that the
dispersion solution corresponds to the average of our selected subsample.
The regridding is performed such that the total flux is conserved.

\subsubsection{Continuum Estimation}
The determination of the continuum is a very tricky problem which is known to
cause difficulties when performing it automatically. For this reason we are not
using the traditional continuum estimates (e.g. spline fitting, local weighting
of polynomials\footnote{e.g. \emph{onedspec}-package in the Image Reduction and
Analysis Facility (IRAF) software package or \emph{norm.pro} from the
Interactive Data Language (IDL) software}) and use a new hybrid method
consisting of the following three approaches:
\begin{enumerate}
  \item fit multiple Gauss model to the data
  \item weight penalty function with variance
  \item iterate 3 times, perform $\kappa-$clipping 
\end{enumerate}

\begin{figure}
\centering
\caption{\textit{Top:} Spectrum with uncertainty. \textit{Upper center:}
Decomposed continuum representation by Gaussians. \textit{Lower center:}
Spectrum with continuum fit and masked regions (ignored when fitting the
continuum). \textit{Bottom:} The extracted feature vectors are solely all pixel
values with a value above/below zero for emission/absorption, all other values
are set to zero.
\label{fig:cont}}
\includegraphics[width=\columnwidth]{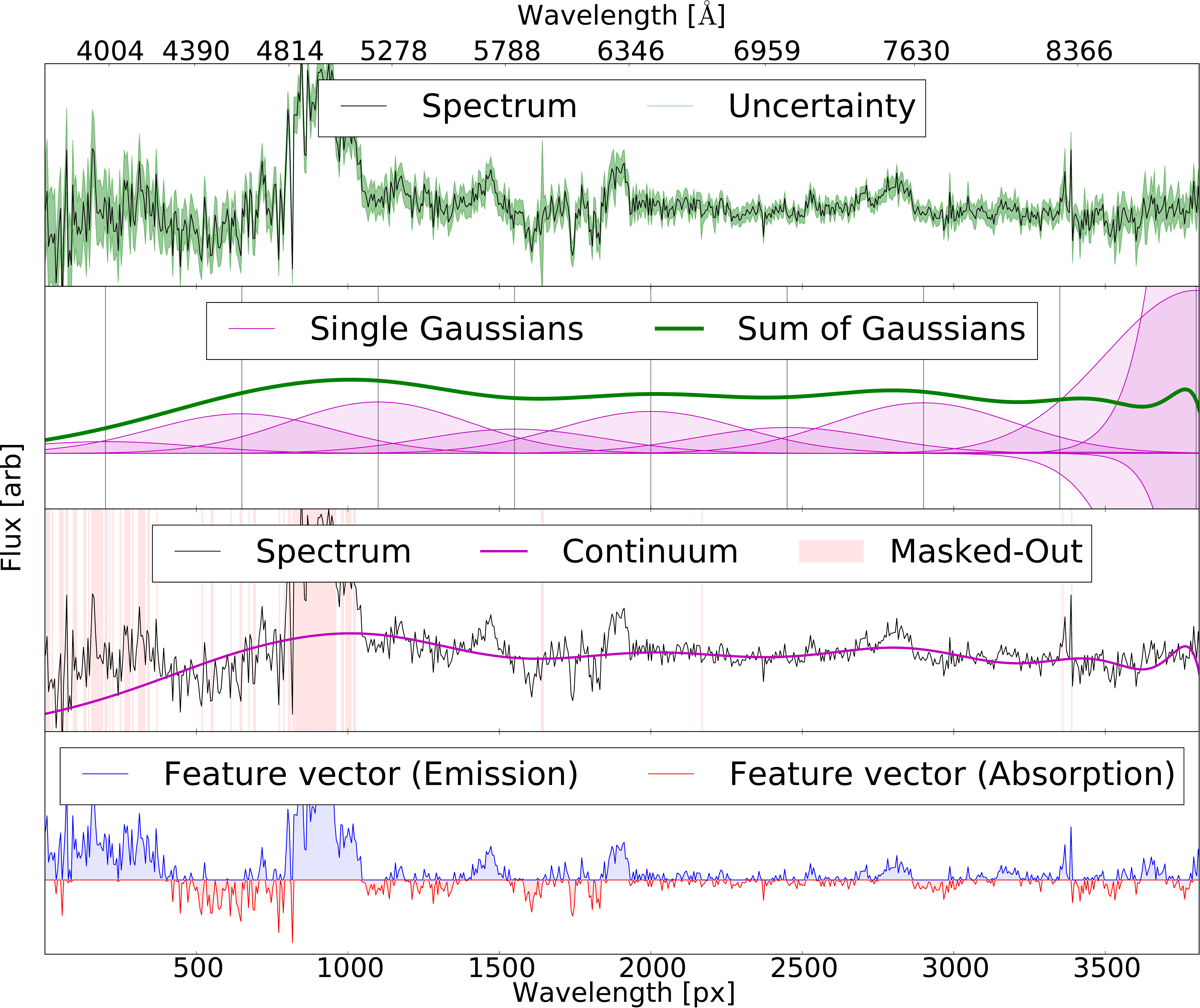}
\end{figure}

In order to save computation time we follow the approach by \citet{Gieseke2011}
and use a multiple Gauss decomposition via gradient based optimization. This
minimizes the risk of over- or underestimating the continuum flux as well as
over-fitting which can be encountered when applying spline fits. 
In order to fit the continuum a number of $n$ normalized Gaussians with the
same width $w\,[\text{px}]$ are placed on the dispersion axis with the
first Gaussian being placed with an offset $\omega\,[\text{px}]$ and all
following with a spacing of $d\,[\text{px}]$. The intensity of every individual
Gaussian is a free parameter to be fitted. In comparison to polynomial and
spline fitting the decomposition is less sensitive to individual spectral
features and the computational effort is significantly lower. An illustration
of the decomposition is shown in \mbox{Figure \ref{fig:cont}}.
Based on the initial fit which is weighted with respect to the uncertainty
$ivar$ (see \ref{subsub:unc}), the root-mean-square is computed. Afterwards
pixel values where the difference between fit and model exceeds $\kappa \cdot
RMS$ are masked out for all future iterations of the continuum estimation. This
is helpful to exclude large-scale deviations and to account for detector or
night-sky artifacts.

Adjacently the spectrum is now normalized with respect to the estimated
continuum $\mathcal{C}$ by a simple min-max-normalization:
\begin{equation}
\mathrm{Flux}_{\mathrm{norm}} =
\frac{\mathrm{Flux}_{\mathrm{raw}} - \mathrm{min}\left(\mathcal{C}\right)}{\mathrm{max}\left(\mathcal{C}\right)-
\mathrm{min}\left(\mathcal{C}\right)}
\end{equation}
such that the continuum of the normalized
flux is located between 0 and 1 and the features are normalized with respect to
continuum. As only the features are of interest for the following task the
continuum is subtracted such that a flat spectrum is obtained. While
testing different pre-processing parameters it turned out that the quality of
the overall redshift is only marginally depending on the parameters
used for estimating the continuum. An overview of the parameters and their
impact on computation time and the quality is given in table
\ref{tab:parTab}. {\bf In contrast to the literature we treat the
Ca-break also as a feature and thus if the continuum behaves smoothly
around the break it can be seen as two close by absorption lines afterwards.}
\begin{table}
\centering
\caption{Parameters used in pre-processing with tested value range and impact on
the outcoming distribution as well as on the time effort for the
pre-processing.\label{tab:parTab}}
\begin{tabular*}{\columnwidth}{cc|c|c}
 & \bf{parameter} & \bf{range} &
\bf{impact} \tabularnewline
 & \bf{description} & \bf{[used value]} & \bf{[result / time]}\tabularnewline
\hline
\hline
\multirow{2}{*}{$n$} & \multirow{2}{*}{number of Gaussians} & 8 - 20 &
\multirow{2}{*}{low / linear}
\tabularnewline 
 & & \bf{[12]} & \tabularnewline
\hline
\multirow{2}{*}{$d$} & spacing between & 300-700 & \multirow{2}{*}{low / none}
\tabularnewline
 & centers & \bf{[450]} & \tabularnewline
\hline
\multirow{2}{*}{$\omega$} & initial center offset & 100 - 400 &
\multirow{2}{*}{none / none}
\tabularnewline
 &  of first Gaussian & \bf{[200]} & \tabularnewline
\hline
\multirow{2}{*}{$w$} & \multirow{2}{*}{Gaussian width} & 100 - 1,000 &
\multirow{2}{*}{low / none} \tabularnewline & &  \bf{[300]} & \tabularnewline
\hline
\multirow{2}{*}{$i$} & number of iterations& 1 - 3 & \multirow{2}{*}{none /
linear}
\tabularnewline
 &  for sigma clipping & \bf{[3]} & \tabularnewline
\hline
\multirow{2}{*}{$\kappa$}	& noise deviation for & 0.1 - 3 
&  \multirow{2}{*}{low  / none} \tabularnewline
 & feature refitting & \bf{[0.3]} & \tabularnewline
\end{tabular*}
\end{table}

\subsubsection{Uncertainties}
\label{subsub:unc}
The SDSS spectra are affected by several uncertainties originating from the
night sky, detector deficiency and read-out noise which are quantified
pixel-wise by the inverse variance $ivar$ which corresponds to the
noise uncertainty $\sigma$ given by
\begin{equation} \sigma = \frac{1}{\sqrt{ivar}}\end{equation}
After re-normalizing $ivar$ with respect to the continuum as described above,
the extracted signal-to-continuum spectra are divided by $3\sigma$ in order to
normalize the noise to values between -1 and 1, those will be called normalized
S/N spectra (NSN-spectra hereafter).
As a consequence the contrast between real signals and noise is further
increased and artifacts originating from a bad sky subtraction/bad pixel are
heavily suppressed.


\subsection{Feature extraction}
To extract the feature vectors we split the spectra into positive and negative
flux components with respect to the fitted continuum (see bottom plot in Fig.
\ref{fig:cont}).
Thereby we create two feature vectors per spectrum. This simplification allows
to keep the entire redshift-dependent information while no longer being
dependent on the continuum shape.
The separation enables us to compute individual redshifts for absorption and
emission. By extracting subregions of this feature vector we can even obtain
redshift information on single spectral regions.
All values above the continuum ($>$0) are included in the feature vector
for emission, all values below the continuum are simply set to zero, the same
holds for absorption, vice-versa. Those extracted vectors are the input for
our $k$NN-search described in Eq.\ref{eq:median}.

\section{Experiments}
\label{experiments}
We conducted two experiments with a different selection of features and
a different reference sample, respectively. In the following they will be named
Experiment 1 and 2.
\subsection{Description of experiments}
Both runs have been done on the full set of NSN-spectra. 
For the first experiment we applied the algorithm to the entire spectra
and just discriminated between absorption/emission. In the second
experiment we limited the dimensionality of the feature vector by just comparing
specific spectral regions where features are expected for the redshift given by
SDSS.

Naively one would expect a high precision in the former method as the full
information content is available and thus the confusion between features of
different origin (e.g. misidentifying $H_{\beta}$ as $H_{\alpha}$) should be
fairly low.
Other emission/absorption signature are available to cross-validate the redshift
and hence minimize the probability of confusion.
On the other hand the obtained global regression value is just valid for
the entire spectrum and thus generalizes the information content too heavily.

\begin{figure}[htb!]
\centering
\caption{Cut-out of the same spectral region for 3
spectra with different redshift. If just this part is available the 
H$_{\beta}$ line is indistinguishable from any of the 
two [OIII]-lines.\label{fig:confusion}}
\includegraphics[width=\columnwidth]{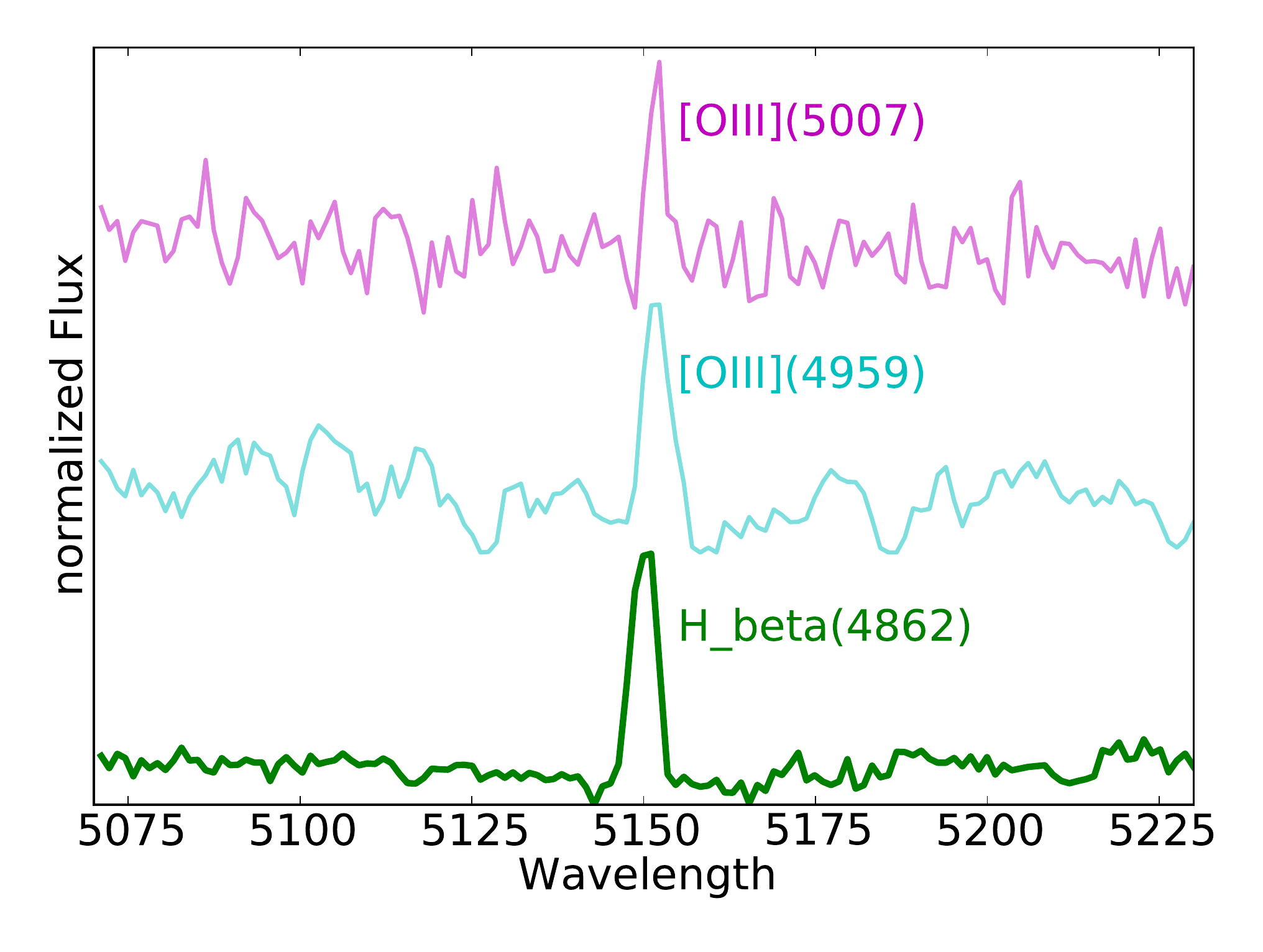}
\end{figure}

For this reason a second experiment was conducted with a comparison restricted
to single regions where prominent emission/absorption signatures are expected.
It is worth noting that this experiment is tailored for detecting shifts of
individual spectral lines. Additionally the methodology can be easily extended
to allow a clustering/classification of the individual lines.
We assume that the redshift of SDSS is correct for the entire spectrum but we
search for redshift deviations of individual components. Since confusion will
have a significant impact on the determination of the redshift we restrict the
redshift deviation of the reference sample to a spectral window $W$ defined
by
\begin{equation}
W=\lambda_1-\lambda_0=\left(1+z_\text{high}\right)\cdot
\lambda_\text{high} - \left(1+z_\text{low}\right)\cdot \lambda_\text{low}
\end{equation}
\begin{center} with \end{center}
\begin{equation}
\begin{split}
z_\text{low} = z_\text{target}- f\cdot
\left(1+z_\text{target}\right)\\
z_\text{high} = z_\text{target}+ f\cdot
\left(1+z_\text{target}\right)
\end{split}
\end{equation}
where $f$ is the allowed deviation from the SDSS redshift
($z_\text{target}$) in units of the speed of light. A list of the spectral
regions that have been taken into account can be found in \mbox{Table
\ref{tab:featTab}}. This list contains lines which are usually strong in
star-forming/bursting galaxies and QSOs. The free parameter $f$ influences the
computational efforts, the chance of confusion (improving for small $f$) and
the sensitivity to outliers where huge redshift deviations were achieved with
large $f$, respectively.
Throughout this paper we will use a value of $f=0.05$.
The big disadvantage of the second experiment is that confusion becomes a
major issue. Especially for the entire data set it might be wise not
to compare spectra to spectra of any redshift as it is likely that, e.g. 
the H$_{\beta}$ ($\lambda4861$) line can look similar to the [OIII]
($\lambda5007$) line, see Fig.\ref{fig:confusion}, which obviously would lead
to a wrong regression value.
The benefit of this concept is its huge flexibility. Redshifts can now be
computed for individual regions independently and thereby shifts can be
detected. A more detailed discussion of the trade-off of confusion and
multi-region regression value determination is given in \mbox{Section
\ref{sect:results}}.
 
\begin{table}[htb!]
\centering
\caption{List of regions taken into account.\label{tab:featTab}}
\begin{tabular}{cccc}
Spectral type & $\lambda_\text{low}$ [\AA] & $\lambda_\text{high}$ [\AA] &
Name \tabularnewline
\hline\hline 
\multirow{9}{*}{Emission} & 2,799 & 2,799 & MgII \tabularnewline
 & 3,346 & 3,426 & NeV \tabularnewline
 & 3,727 & 3,729 & [OII] \tabularnewline
 & 3,798 & 3,835 & H$_{\epsilon}$, H$_{\zeta}$ \tabularnewline
 & 4,102 & 4,102 & H$_{\delta}$ \tabularnewline
 & 4,341 & 4,363 & H$_{\gamma}$ \tabularnewline
 & 4,861 & 5,007 & H$_{\beta}$, [OIII] \tabularnewline
 & 6,550 & 6,584 & H$_{\alpha}$, [NII] \tabularnewline
 & 6,716 & 6,731 & [SII] \tabularnewline
 \hline
\multirow{3}{*}{Absorption} & 3,934 & 3,969 & H+K \tabularnewline
 & 5,173 & 5,173 & Mgb \tabularnewline
 & 5,890 & 5,896 & NaD \tabularnewline
\end{tabular}
\end{table}

\subsection{Maximum deviation limit}
One of the prerequisites in using the $k$NN approach was that a clear
separation between noise and signal can be made. In principle there are two ways
to reject spectra with no signal, the pre- and the post-selection. 
To pre-select one assumes that a signal has a certain shape and exceeds a given
S/N limit. This can be simplified further to a measure which compares the
average of a spectral region with a nominal value.
As this pre-selection requires detailed knowledge about the
shape/size/symmetry of spectral features physical knowledge about the
morphology of lines is required.
In order to be independent of physical assumptions\footnote{Obviously the
reference values by SDSS are obtained via physical modelling.} the possibility
of a post-selection is chosen. The selected concept assumes that the
deviation of the redshifts of the nearest neighbors over all targets follows a
smooth distribution. For this distribution an upper limit can be (freely)
selected which separates redshift estimates into good and noisy ones. This
maximum deviation limit will be abbreviated by MDL.

An even bigger advantage of this
method is that it allows to experiment with this free parameter in the
evaluation stage, such that the $k$NN search is not performed for every
individual value of MDL.
\subsection{Validation strategy}
In order to avoid biases in the regression values and when tuning
the parameters the leave-one-out strategy is used. This means that the closest
object (which is always the object itself) is not used for determining the
redshift. 

The fundamental assumption that most SDSS redshifts are correct was already
discussed in subsection \ref{subsect:prereq}.
Assuming now that all redshifts are correct we can compute something like
a completeness, a correctness and a sensitivity. The completeness is a quite
straightforward measure. It is the fraction of objects for which a
redshift could be determined within the respective acceptance limit. In
contrast to that the correctness is the fraction of objects where the
computed and the SDSS redshift agree within their errors. Finally the
sensitivity gives the reliability of all redshifts, i.e. it gives the
typical deviation from the redshift. Therefore the standard deviation of
the difference between SDSS and computed redshift of all valid spectral
features is computed.

\subsection{Parameter tuning}

\begin{figure}
\centering\caption{Normalized completeness, sensitivity and correctness
tested against different values of $k$ and MDL. MDL is chosen to be
0.0015 as marked.
\label{paramEv}}
\includegraphics[width=\columnwidth]{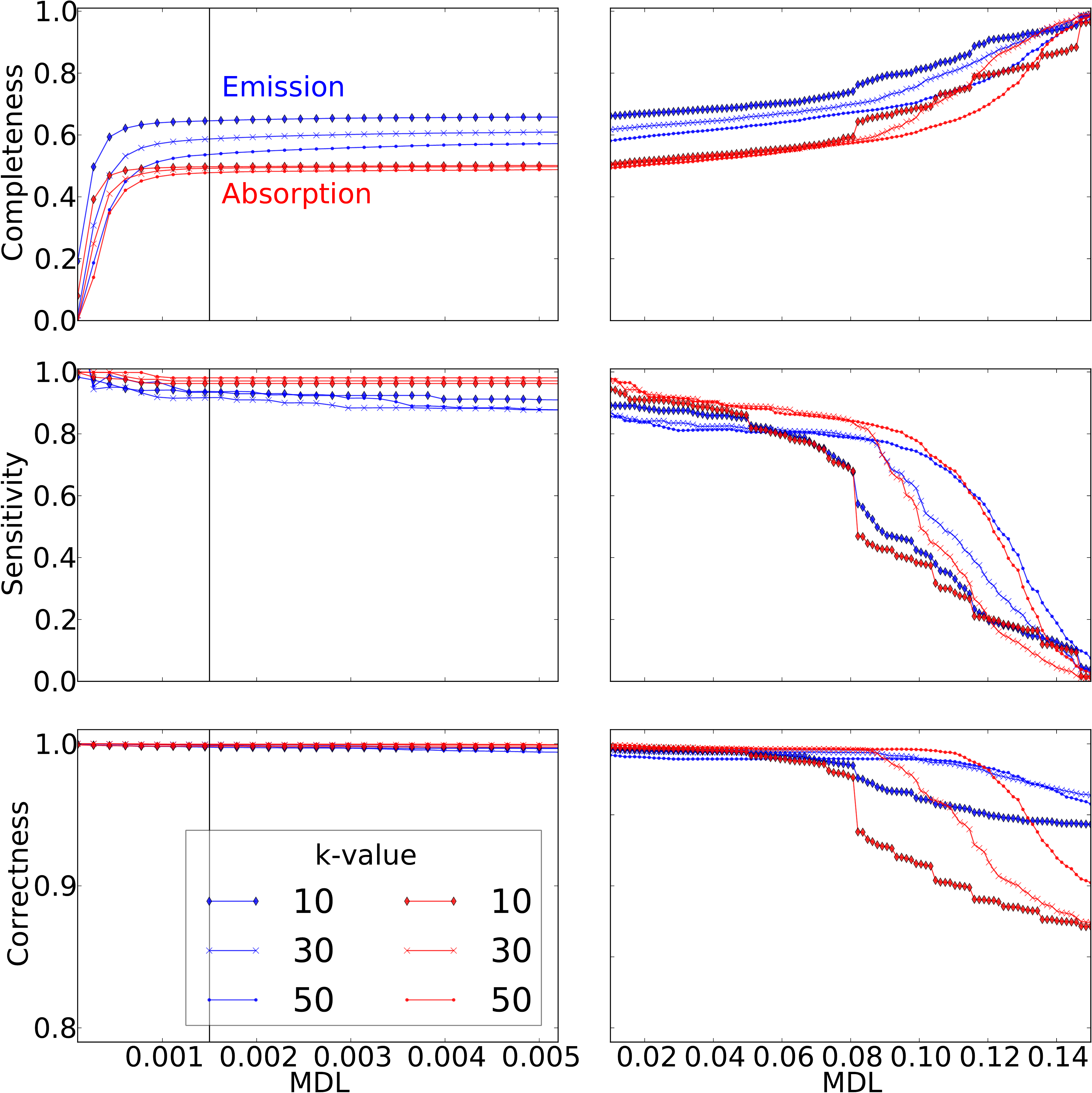}
\end{figure}

Despite the parameters described in the pre-processing step only two parameters
have to be fine-tuned for the regression step\footnote{Note that this is only
partially true because different ways of computing the redshift and
calculating the deviation exist. Besides the parameter tuning one has to
choose a similarity measure and pre-process and select the features,
accordingly.}.
Those are the number of $k$ nearest neighbors used for the comparison and the
MDL which marks a spectrum to be reliable. With the test strategy described in
the previous section, this fine-tuning can be solved on a discrete grid, see
\mbox{Figure \ref{paramEv}}.
In this plot two separate things are shown, the large scale behavior of the
properties on the right side and on the left side a zoom-in to the lowest values
of the MDL.

With increasing MDL, which is equal to accepting more noisy spectral features,
the properties behave just as expected; while the completeness is increasing, the
sensitivity and correctness of the model are decreasing. One can further see
that the completeness is a rather flat function up to a MDL of 0.08 where it
starts a more rapid, step-wise increase. The regression model breaks already
down at a MDL of 0.05 where the sensitivity and correctness show a steep
decrease.
As the increase of the completeness is only very tiny for large values of MDL,
we now focus on the region of very tiny values of MDL.
 
On the small scale the completeness strongly depends on the choice of the MDL
and slight increases of the MDL yield a strong increase in completeness.
Then the behavior becomes very flat and thus the gain by further increasing the
MDL is only marginal (when increasing the MDL 5 times, the completeness fraction
increases by less than 2\% ). It is worth noting that the completeness depends
quite heavily on $k$.
Smaller $k$ values result in a more complete regression model. This
indicates that the number of good references is of the order of 10-20 as for
larger $k$ values apparently more deviation is introduced in the regression
model.

The sensitivity of the regression model is only decreasing in the
beginning and follows then a fairly flat behavior with a slightly decreasing
tendency. The dependence on the number of used neighbors is only marginal,
though one can see that emission is in favor of a low $k$ (little number of
reference objects), while the sensitivity of the redshift in absorption is
slightly better for higher values of $k$.

In the end the fraction of outliers on the good regression side is only
slightly changing with increasing MDL and $k$. The decrease over the entire
tested range is of the order of 0.5\%.

\begin{figure}
\centering\caption{Dependence of test properties on $k$. For
this plot MDL is fixed to 0.0015. $k$ is chosen to be
40 as marked.\label{kDep}}
\includegraphics[width=\columnwidth]{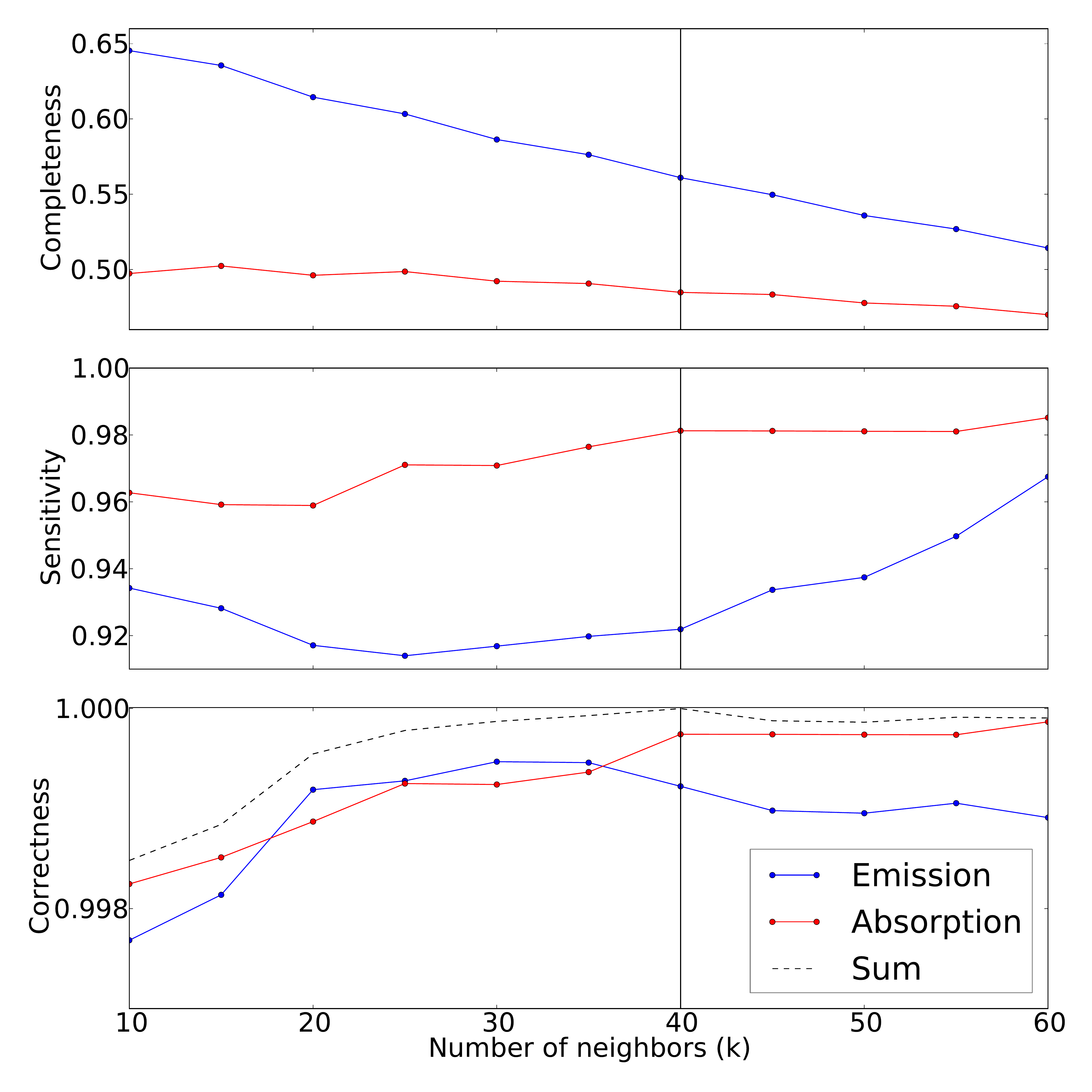}
\end{figure}

The flat increase in completeness for MDL values larger than 0.001 allows
us to minimize the effects on the sensitivity and correctness. 
We analyzed the impact of the choice of $k$ on the different testing properties
as well. The behavior of those with a fixed value of MDL of $\approx0.0015$ can
be seen in \mbox{Figure \ref{kDep}}. An increasing number of nearest
neighbors improves the sensitivity at the cost of a lower completeness. Thus as
for the MDL the choice of $k$ depends strongly on the desired completeness and
precision.

\subsection{Computational efforts}
Applying the method described above to the test set is already quite time
consuming on a single machine. It is evident that the computational effort for
\mbox{3$\,$Million} spectra is multiple times larger than with 16,000, i.e. the
time complexity of a brute force $k$NN search scales with O(n$^2$), thus the
calculation time would already be of the order of years on a single machine.
For future surveys this number will increase even faster such that more
efficient approaches have to be found to resolve that problem. To speed up the
calculation we parallelized the computation of the distances. The results
presented here should solely give an overview on what is possible with even the
simplest methods when such a huge data amount is available.

It is worth noting, that an online nearest neighbor search of incoming
data (streaming) with a spectral database of the size of SDSS ($\approx$
3,000,000 spectra) is computational easily feasible on a modern laptop.
Assuming that a new instrument \citep[e.g. 4MOST, ][]{2012SPIE.8446E..0TD} 
will obtain 2,400 spectra simultaneously, the approximate comparison time
is of the order of 40h/core using a standard Python implementation.
Using a machine with a simple GPU and a C-implementation will yield a
speed-up of at least 100 compared to the single-CPU machine and can evaluate
such a huge amount of data ($<$30min) in less than the typical exposure time.
Fortunately the computation of the distances can be perfectly parallelized,
hence the presented method is well suited for scaling to larger surveys on
modern computer architecture.

As already stated the computational effort is also strongly depending on the
number of reference objects used for comparison to and obviously that is one
of the most important screws to tune in order to minimize computing time. On
the other hand the impact of selection effects is minimized by increasing the
number of reference objects which have to be chosen in the most unbiased manner.

%% file: Results.tex
\section{Results}
\label{sect:results}
In the following we will use the median absolut deviation (MAD) as
deviation measure, which is defined as 
$$\text{MAD}_{\overline{v}} =
\text{median}\left(\lvert\overline{v}-\text{median}\left(\overline{v}\right)\rvert\right)$$
for a list of values $\overline{v}$. {\bf In the following we will always use
the normalized difference in redshift which is defined as:
$$\Delta Z_{\text{norm}} =
\frac{z_{\text{kNN}}-z_{\text{SDSS}}}{1+z_{\text{SDSS}}}$$ which corresponds to
the difference in velcocity in terms of c in the rest-frame of the SDSS
redshift.} In Fig.
\ref{paramEv} one can see the behavior of the completeness, sensitivity and
correctness as a function of the MDL as well as for different $k$.
The curves follow the expected behavior; decreasing MDL
will yield a low completeness but therefore high-quality redshifts.
In the middle is a plateau until the MDL exceeds $\approx$0.10. 
Beyond this value the completeness starts to converge against 1 and the quality
of the redshifts against 0. While for the value added catalog a high
completeness is desirable under moderate loss of sensitivity, thus MDL$=$0.07
and $k=$10 are chosen. This increases the fraction of objects with a reliable
redshift either in emission or in absorption up to a total of 80\%.
With this choice of parameters we have still a better
sensitivity than SDSS with a significantly lower value in completeness (in SDSS
$\approx$96\% of the targets have NO redshift warning).
As stated earlier the choice of the reference sample, especially at high
redshifts, will increase this fraction of our method significantly, just when
excluding spectra with $z>0.25$ the completeness increases to $\approx$ 90\%.
 
In the following we concentrate on the detection and verification of
outliers using MDL$=0.015$ and $k=40$. With that choice we have traded a 
high sensitivity for a lower completeness of $\geq$50\%.
This enables us to efficiently detect outliers which show a wrong or
multiple redshift components.
In the following the outlier detection for both experiments will be discussed in
detail.
\subsection{Experiment 1}
When using the entire spectral range for computing the redshift, we can obtain
redshifts for 56\% (emission), 49\% (absorption) of the spectra.
In \mbox{Figure \ref{fig:emPars}} the evaluation of the achieved performance is
presented.
In the second row of each figure one can see the frequency of
deviations for emission and absorption, respectively\footnote{Note that the bin
width is changing by a factor of 25 from the left to the right side. For this
reason the frequency between the two plots is not directly comparable}.
As expected there is an exponential drop-off and a underlying uniform contribution.
The top figure shows respectively the relative deviation (in units of the speed
of light) between the redshift by SDSS and the computed ones. For nearly all of
the objects with prominent features this deviation is below 0.1\%$\,$c which
corresponds roughly to the SDSS resolution.

In emission one can see three groups of outliers, three points between a 
redshift of $0.2\leq z \leq 0.3$ (G1, magenta background), a straight line
in the lower right of the plot (G2, cyan background), and three points
that significantly deviate from the expected redshift below a redshift of $z\leq
0.1$ (G3, blue dots).
The cause for each of the outliers groups is different but anyway well
understood. The members of G1 are affected by the lack of
reference objects in a comparable redshift range ($z>0.2$) which agrees
perfectly with the distribution shown in \mbox{Figure \ref{fig:redDist}}. Thus
the nearest neighbors will all have a lower redshift, moving all of those points to
this region in the plot. It is worth noting that naively one would expect all of
those points to lie on a horizontal line as well as the deviation from the
reference set should for all objects be the same. In fact it turned out
that the lowest point in this group is a truly shifted object. G2 is
actually a superposition of the problem just described and what was defined
earlier as confusion. The confusion occurs since the relative shift in redshift
of $\Delta Z_{\text{norm}}\approx-0.25$ corresponds roughly to the shifts between
$H_{\alpha}-H_{\beta}$ ($\Delta Z_{\text{norm}}=0.26$), $H_{\alpha}-[OIII]$
($\Delta Z_{\text{norm}}=0.24$), $[NII]-H_{\beta}$ ($\Delta
Z_{\text{norm}}=0.26$) and $[NII]-[OIII]$ ($\Delta Z_{\text{norm}}=0.24$). In
this case the spectra usually show strong emission in either $H_{\beta}$ or
[OIII] which are then (due to missing references) misidentified as [NII] or
$H_{\alpha}$.
Finally spectra with
real shifts are likely to be observed close to the horizontal green line.
They are further discussed in \mbox{Subsection \ref{sub:single}}. The
behavior of the noisy features can be explained by another superposition of two
effects. The first group of objects is the one where the relative deviation is
fairly low over the entire redshift range.
Those objects are the result of the choice of the MDL - their redshift is still
very accurate but they were moved to the uncertain features.
A large number of spectra can described very nicely with the applied model. This 
indicates that the MDL was selected quite conservatively. The rest of the data
points in this plot do not show any signal of an emission feature thus they are
just a random selection of redshifts from the initial distribution shown in
\mbox{Figure \ref{fig:redDist}}. The distribution of redshifts is fairly well
approximated by a Gaussian (mean=0.14 and standard deviation=0.10). The
functional form (cf. blue background plot in upper row) is:
\begin{equation}
\left(\left(0.14\pm0.1\right)-z_{SDSS}\right)/\left(1+z_{SDSS}\right)
\label{eq:blueRegion}
\end{equation}

In absorption two outliers could be detected which show some anomalies
that are well described by the computed redshift. Even redshifts with high 
MAD are still fairly reliable, supporting the restrictive limit
on the MDL. The precision in absorption is of the same order as 
the emission, one per mil in units of the speed of light. Obviously the
chance of confusion is dramatically smaller than for the emission which is the
consequence of the lower number of potential features. Typically in a regular 
galaxy only three strong absorption features can be observed.

\subsection{Experiment 2}
In contrast to the first experiment the number of potential nearest neighbors of
a specific spectral region depends now strongly on the choice of the redshift bin
and additionally on the likelihood of the respective feature appearing in a
galaxy spectra. This makes it inevitable to discuss the chosen regions
individually. To have still a good comparison of the redshifts between
the different regions, the MDL is set to 0.0015. For the sake of completeness
all the figures comparing the noisy and the good features are presented in the
\mbox{Appendix \ref{sec:app}}. Without restricting the results any further the
number of potential outliers increased drastically due to the problem of
additional confusion with different spectral features as well as to the limited
number of used reference objects. Thus in order to remain clear 
and minimize the effect of methodological artifacts the deviation/outlier
constraint is not just tested for $k$=40 but for an entire list of nearest
neighbors, namely $k$=[5,10,20,30]. 
If the MAD violates the MDL or if the
computed redshift agrees in its tolerance with the SDSS redshift for any $k$,
the object is not marked as an outlier. Additionally objects which have
redshifts $z<0.05$ or $z>0.3$ are automatically excluded from the outlier
detection algorithm as here the limited number of comparison objects introduces
spurious redshifts. As the different regions are biased by different effects
they are adjacently discussed in more detail.

\afterpage{%
 \clearpage 
 \onecolumn

 \begin{figure}
 \caption{The evaluation of the performance for emission (a) and
 absorption (b): The relative deviation from the SDSS redshift as function of
 the SDSS redshift (top), the distribution of the MAD of the calculated redshift
 (middle) and the frequency of the relative deviation (bottom) are shown for
 the good (left) and the rejected noisy (right) spectral features, respectively.
 The blue background shade in the upper right figure reflects the objects which are
 entirely dominated by noise and thus their computed redshift just reflects a
 random draw of redshifts from the initial distribution, see
 \mbox{Equation \ref{eq:blueRegion}}.\label{fig:emPars}}
  \centering 
 \subfigure[Emission]{\includegraphics[width=0.94\textwidth]{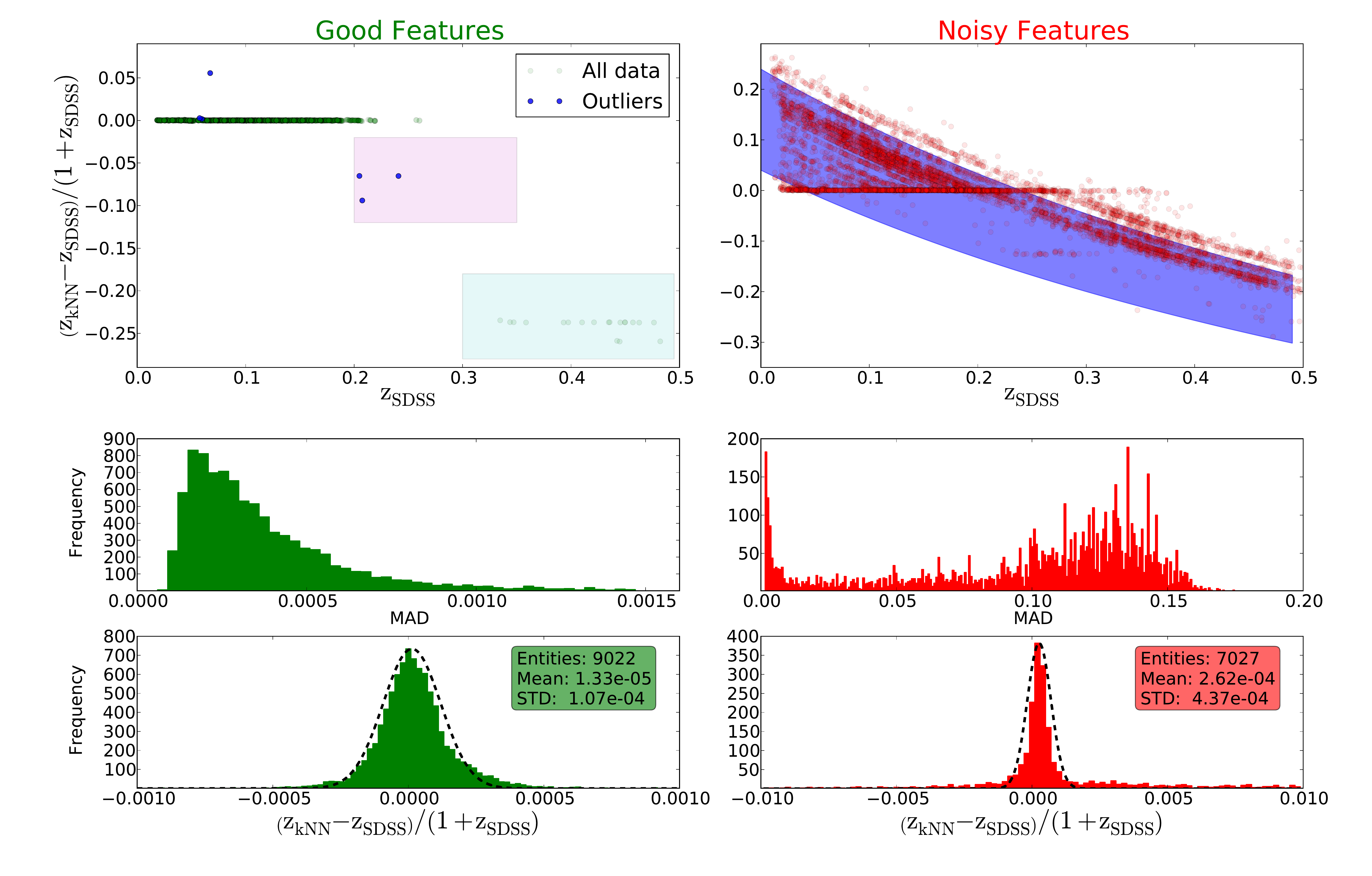}}\hfill
 \subfigure[Absorption]{\includegraphics[width=0.94\textwidth]{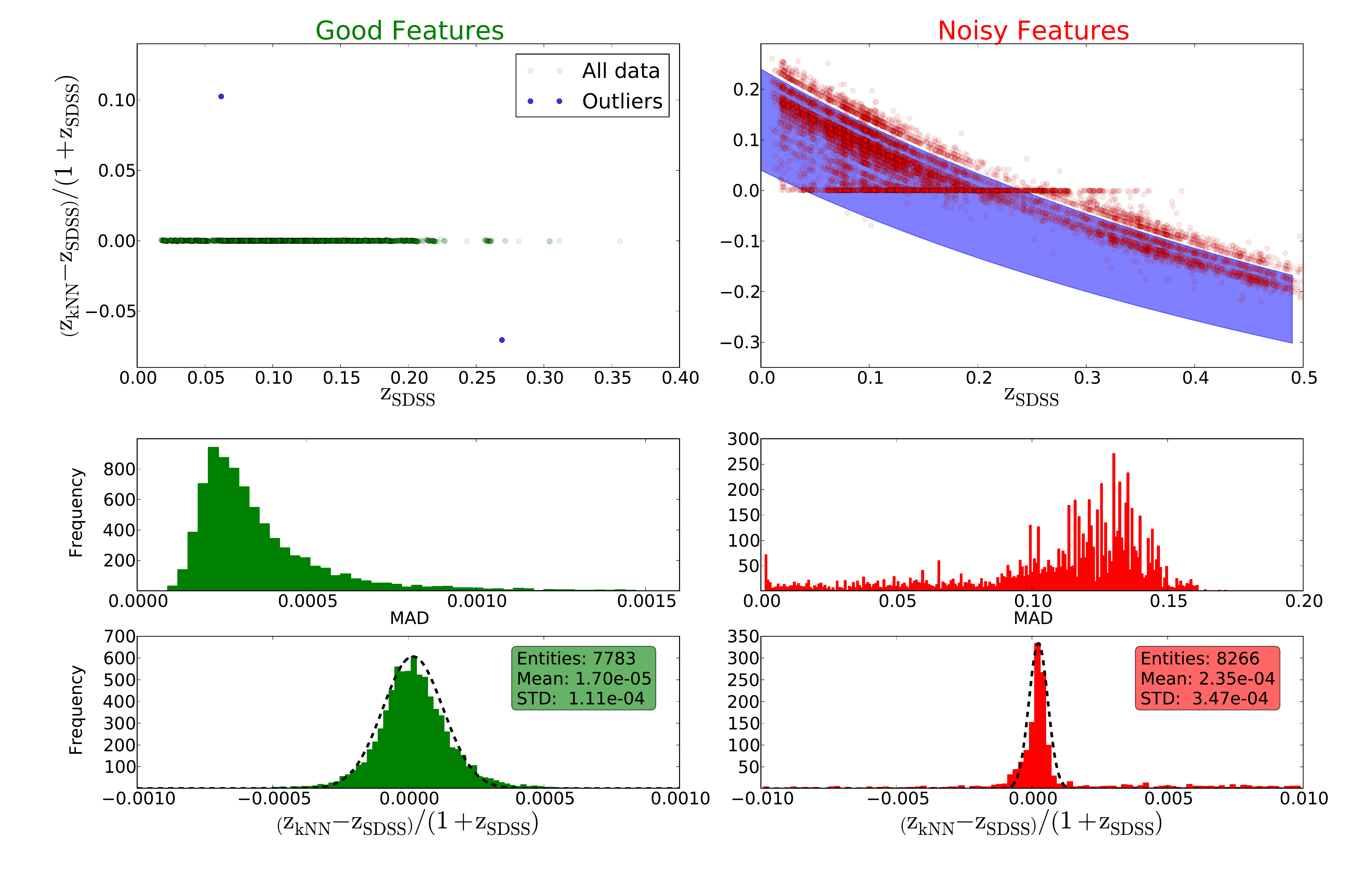}}
 \end{figure}
	\ifreftype 
		\onecolumn
	\else 
		\twocolumn
	\fi
}

In the following we discuss the individual spectral emission and absorption
features. Additionally groups and individual outlying spectra are discussed.
{\bf Exemplarily extensive plots for two spectral features are shown in the 
\mbox{Appendix \ref{pl1},\ref{pl2}} for H$_{\beta}$ and NaD respectively.}
\subsubsection{Emission}
\paragraph{MgII, NeV ($\lambda$2,799, $\lambda$3,346-3,426)} For those
spectral regions a redshift of $z=$0.45/0.18 is required to allow for a
redshift determination. As the number density of objects is fairly sparse for
such high redshifts and additionally the NeV feature does not occur in many of
those spectra, none of the redshifts can be trusted.
\paragraph{[OII] ($\lambda$3,727-3,729)} This feature does not occur in all
star-forming/active galaxies such that less than half of the redshifts could be trusted.
Even in this small fraction of objects two outliers were detected which both
show an actually shifted [OII]-line which is correctly described by the value
determined by us.
\paragraph{H$_{\epsilon}$, H$_{\zeta}$ ($\lambda$3,798-3,835)} One of the
objects found to have a shift in the [OII]-feature could be rediscovered. Both other additional
spectral features are real.
\paragraph{H$_{\delta}$ ($\lambda$4,102)} This spectral feature just appears
in emission for star-forming/bursting and active galaxies. The number of reference
objects exhibiting a clear sign of emission is fairly rare.
In the corresponding plot one can see that two straight regions are apparent at
$\Delta Z_{\text{norm}}=0.04/-0.04$ which are caused by confusion. The remaining
object shows some very strong noise in the vicinity of the expected spectral 
feature.
\paragraph{H$_{\gamma}$,[OIII] ($\lambda4,342-4,363$)} 
The only two remaining spectra have a $\Delta Z_{\text{norm}}=-0.032$. When
investigating the origin of this shift, it appears that the shift is dominated
by noise as the number of active and starburst objects (objects which possibly
emit strong Balmer lines) in the specific redshift bins is very low ($<$5). So
when selecting the redshift those few objects are strongly dominated by noise. 
Consequently this feature is not very reliable as long as not a
representative reference sample can be selected.
\paragraph{H$_{\beta}$, [OIII] ($\lambda$4,861-5,007)} In this spectral region
the impact of confusion becomes dominant. 14 objects show a reasonably low deviation
to be marked as good estimates. 
The horizontal line at $\Delta Z_{\text{norm}}=0.03$ is caused by a
misidentification of the red [OIII]-line with the $H_{\beta}$-feature. The line
at $\Delta Z_{\text{norm}}=0.01$ is due to the confusion between the red and the
blue [OIII]-line. The negative confusion at $\Delta Z_{\text{norm}}=-0.03$ is
the reverse effect of the first one. Another horizontal component at $\Delta
Z_{\text{norm}}=-0.055$ is caused by a misidentification between the blue
[OIII]-line and HeII emission at $\lambda$4,685.

Apart from all this confusion there is one regular shift which cannot be
confirmed due to the lack of other emission features. The MAD for
this object (0.0014) is close to the MDL so a lower choice of the MDL would
tag this object as unreliable.

\paragraph{H$_{\alpha}$, [NII] ($\lambda$6,550-6,584)} The outlier on the very
top of the plot was already marked by the first run and is a truly shifted spectral feature. One
of the shifts of the remaining two outliers is the result of an
$H_{\alpha}$-line in absorption and emission such that the red [NII]-line was
mistaken for it and in the other a very weak [NII] emission line led to
confusion with the $H_{\alpha}$-line.

\paragraph{[SII] ($\lambda$6,716-6,731)} The only object marked in the plot
was also detected in the $H_{\alpha}$-line as an outlier. It was already marked as an outlier in
Experiment 1.

\subsubsection{Absorption}
\paragraph{CaII (HK) ($\lambda$3,934-3,969)}All three targets highlighted as
outlier are all truly shifted spectral features, one of them being the object
already detected in emission (cf. \mbox{Experiment 1}).
\paragraph{Mgb ($\lambda$5,173)} 
Six of the objects are located on a horizontal line around \mbox{$\Delta
Z_{\text{norm}}=-0.06$}.
This corresponds to a misidentification of the Mgb absorption with the $H_{\beta}$
in absorption. Indeed all highlighted objects show a very prominent
$H_{\beta}$ feature in absorption. Two of the remaining objects have a very
strong absorption feature originating from deficient nightsky subtraction, not
properly described by $ivar$. Three objects are active galaxies and show
extremely strong emission features in this region. The number of active
galaxies in the reference sample is not sufficient to reproduce this behavior.
The remaining object shows a true shift in the Mgb line.
\paragraph{NaD ($\lambda$5,890-5,896)} For seven spectra a shift of the NaD
could be confirmed by a manual inspection, for all the others a badly
subtracted sky at around $\lambda$7,200 was not described correctly by $ivar$,
leading to a very prominent absorption feature which was mistaken for NaD.

\subsection{Manually Investigated Objects}
\label{sub:single}
To validate the method a manual inspection of the outliers is
mandatory. A spectrum was investigated if it was selected as an outlier in
any of the spectral regions (from Experiment 1 and 2) and if it was not part of
one of the horizontal lines introduced by confusion. The outliers have different
origins which can be roughly classified into three groups: objects with real
multiple redshift components (true), objects with detector/nightsky artifacts
which were not properly described by $ivar$ (fake) and objects where the
redshift computation simply failed (wrong).

37 objects were eventually investigated manually, three of those have been
marked by several features as outlier. 38\% (14) of the outliers are spectra
with truly shifted redshift components. In 11 of those the shift between the
redshift components is lower than $10,000\,$km/s thus those components do
certainly have a physical origin. The remaining three spectra of the true class
are likely to be superpositions and/or lensed objects. The fake category contains
10 objects where a badly described detector/nightsky artifact was confused with
NaD or Mgb absorption. It is impossible to exclude those objects
previously as there is no unique position/indication of the existence of such a feature. The
13 spectra in the wrong class are mainly a result of a biased reference
sample which additionally contains a low number of active and star-bursting
galaxies. There is a good chance that the fraction of those objects can be
significantly decreased if a more representative reference sample is used for
the comparison.

A short summary of all manually investigated objects with identifier, SDSS and
computed redshift can be found in \mbox{Table \ref{objSummary}}.

\subsection{Most Prominent Outliers}
\label{sub:outliers}
\begin{figure}[ht!]
\centering
\caption{The three most extreme outliers obtained from
our regression model are shown. The green line in the
background is the SDSS spectrum with the gray spectrum at
the bottom being the typical noise deviation. The red
curve shows the fitted spectrum with a redshift as
obtained by SDSS, the blue curve is the overplotted
spectrum with the redshift as obtained by our regression
model.\label{fig:specialOnes}}
\includegraphics[width=0.5\textwidth]{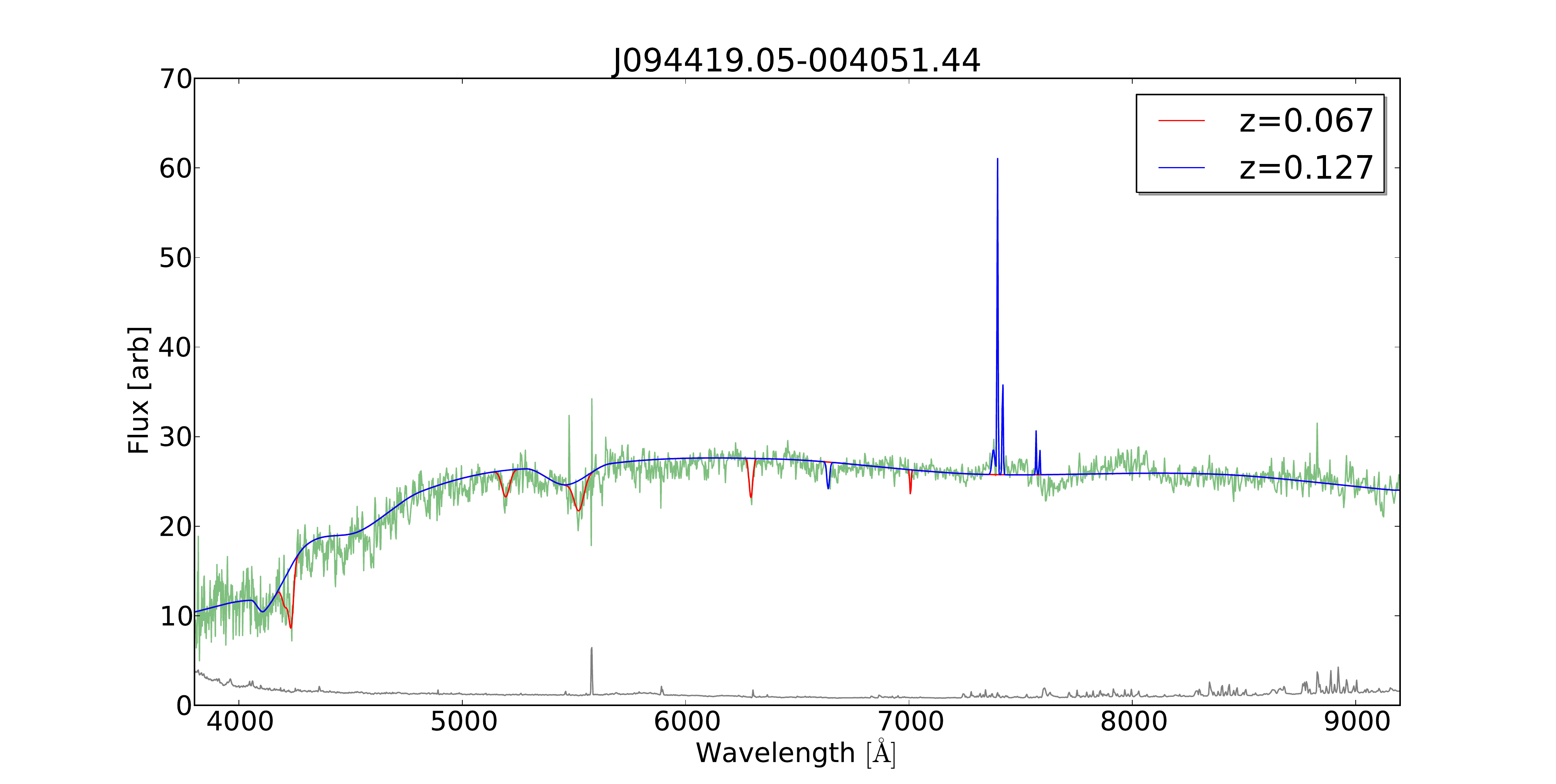}
\includegraphics[width=0.5\textwidth]{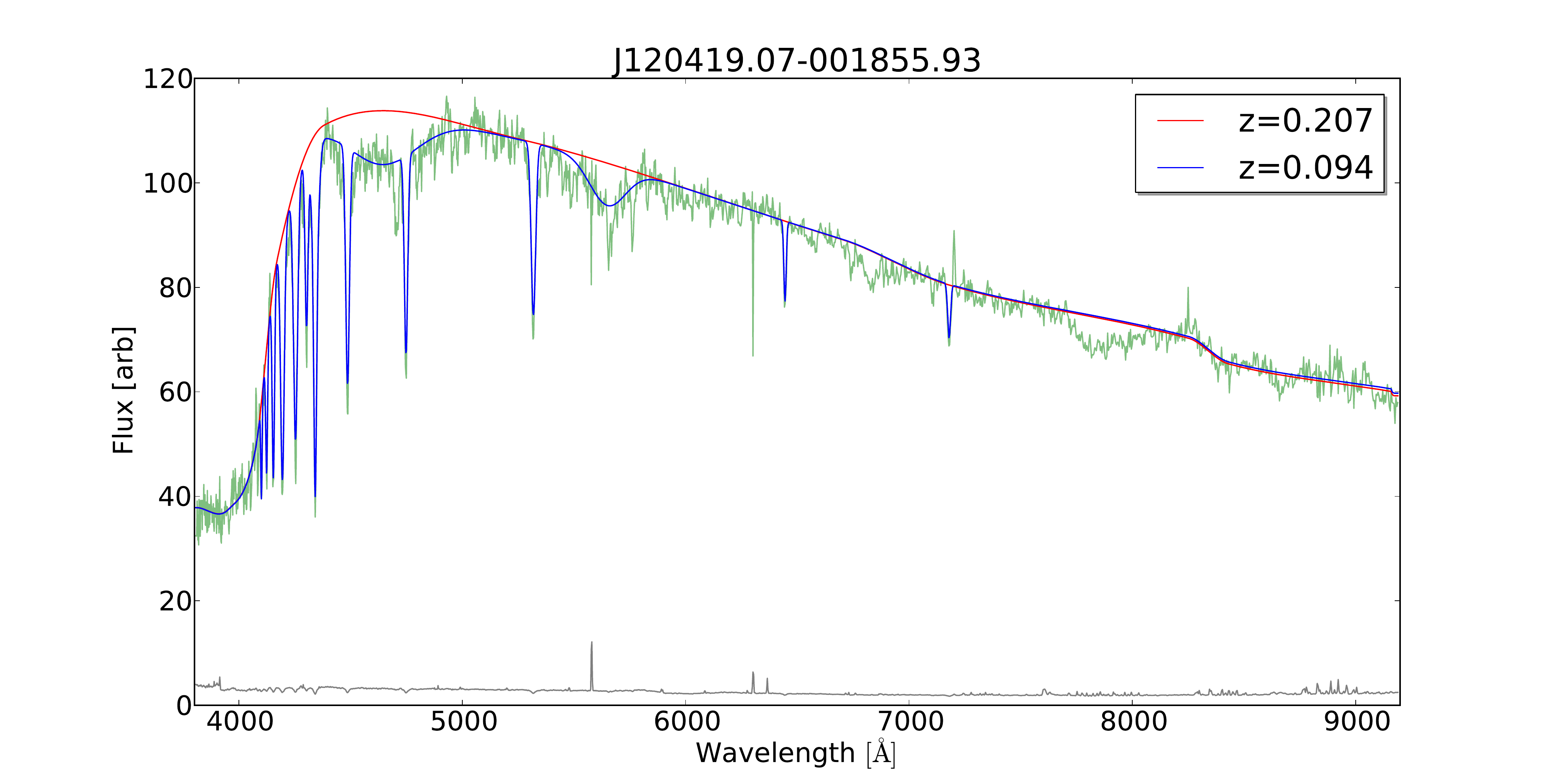}
\includegraphics[width=0.5\textwidth]{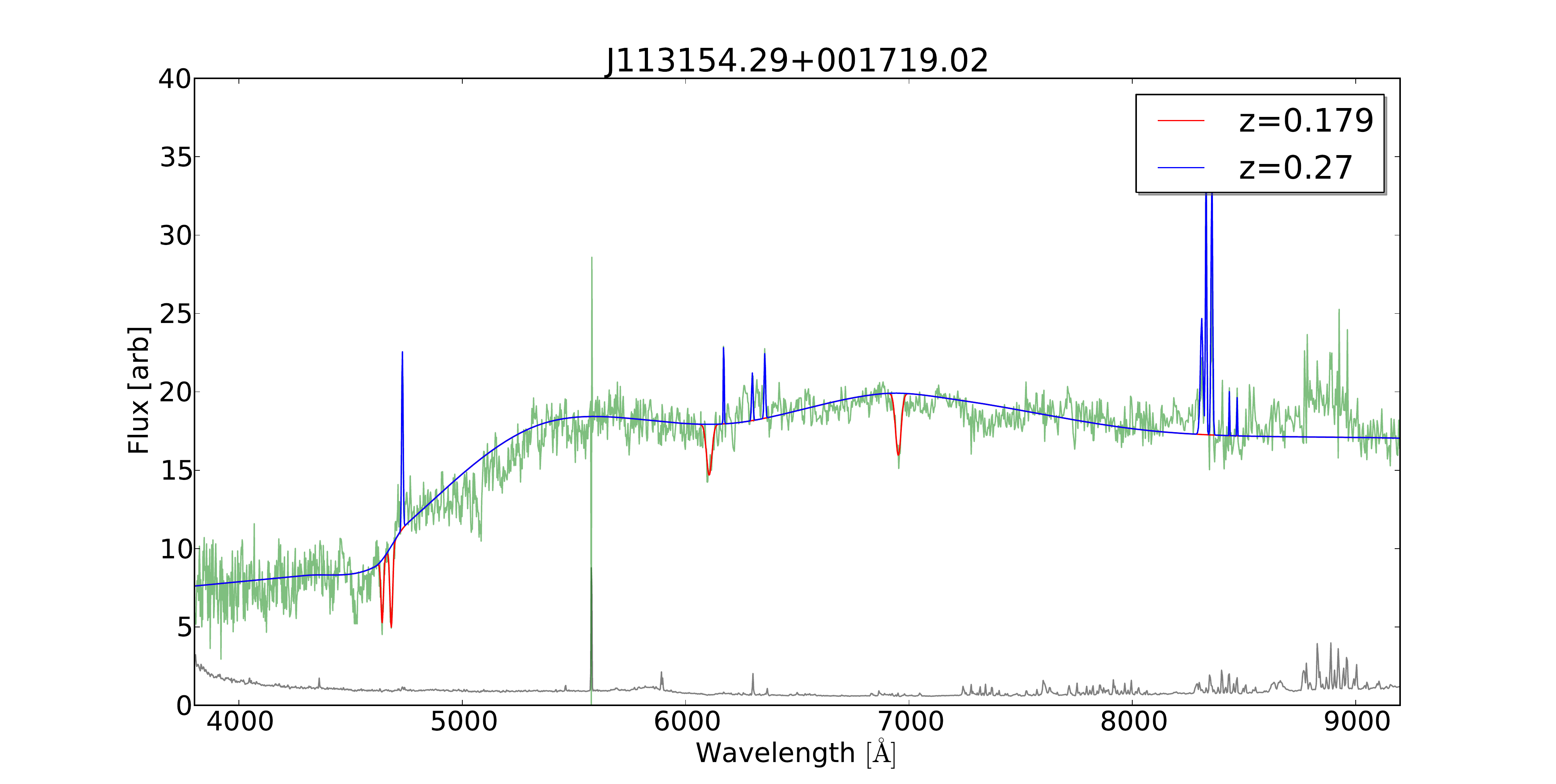}
\end{figure}

\begin{figure}[ht!]
\centering
\caption{SDSS i-band image of J094419.05-004051.44 (smoothed with Gaussian blur
of 3 pixel width). In the right image the asymmetric arc has been overplotted by
a green circle.\label{fig:arc}}
\includegraphics[width=0.45\textwidth]{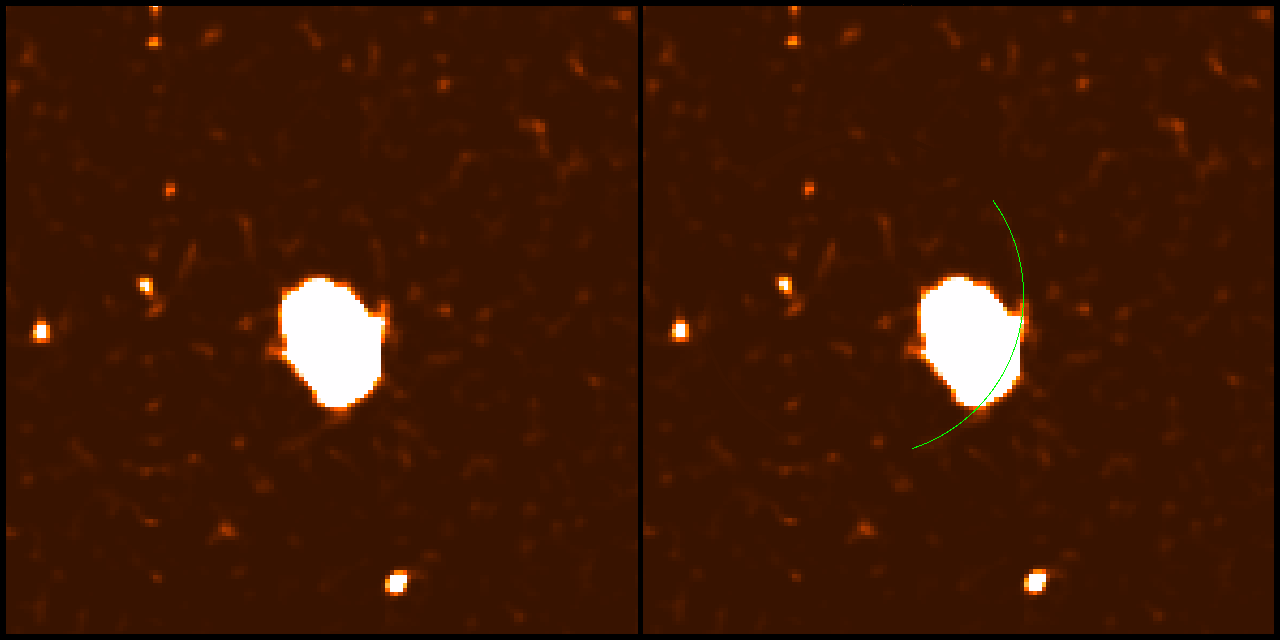}
\end{figure}

The most prominent outliers will be shortly described here to emphasize the
power of this outlier detection scheme. In \mbox{Figure \ref{fig:specialOnes}}
one can see the three truly shifted objects with the highest velocity offset.
While the first two (J094419.05-004051.44, J120419.07-001855.93) were tagged
even independently by the separate runs, the last one (J113154.29+001719.02) did
not show up in the second experiment as the relative shift between our computed
and the SDSS redshift (0.077$\,$c) exceeds the allowed range of the shift 
(0.060$\,$c).
In the first and last object the model applied by SDSS describes the absorption
behavior quite well but the emission features are not described at all such
that a second component with a strongly shifted redshift is needed to describe
those.
While they quite nicely demonstrate the power of the method these objects are
astronomically less interesting. It is likely, due to missing signs of
interaction, that those are just simple superpositions of objects. In the
$i$-band of the first object a tiny and asymmetric arc 
(cf. Fig. \ref{fig:arc})
can be seen which could indicate a lensed object.
The redshift of the second object was estimated entirely wrong by SDSS as
apparently none of the template models was able to describe the continuum and
the line behavior at the same time. The newly estimated redshift on the other
hand describes the spectrum quite well. While the new fit does not support the
existence of another component it is worth noting that on the SDSS image a clear
symmetric arc can be seen at a distance of a few arc-seconds.

\subsection{Summary Of Outliers}
In \mbox{Table \ref{sec:app}} all outliers found are summarized.
Targets where a true feature exists are marked. The possible origins for the
existence of multiple redshift components are miscellaneous. For a very high shift
between the redshifts the most likely explanation is a chance
superposition of two objects.
If those are at an according distance to each other and to the observer an arc
due to gravitational lensing might be observed. The number of gravitational
lenses in the near universe is very limited so far \citep{1998Ap&SS.263...51M}. 
Spectra with the velocity shifts
between the lines lower than $<10,000\,$km/s might be good candidates for being
super-massive black hole binaries (SMBHB,
\citealp{2011ApJ...738...20T,2012ApJ...745...67F, 2012NewAR..56...74P}). 
The kinematics of the broad line region are a very common cause for such
observed line shifts as well \citep{2011ApJS..194...45S}.

Eventually one could only discriminate between the different origins by either
deep-imaging (lenses) or by follow-up spectroscopy (SMBHB,
\citealp{2014ApJ...789..140L}). High-resolution imaging in the multiple
wavelengths could also distinguish single from multiple sources
\citep[e.g.][]{2009ApJ...697...37R}.

%% file: Discussion.tex
\section{Summary}
This paper presents a new methodology which performs a redshift computation
based on pre-exisiting SDSS redshifts. The aim is to obtain improved redshifts
for emission and absorption as well as for individual spectral features. This
enables astronomers to detect spectra with multiple redshift components. The
basic principle of the presented method is to perform a self-consistency check
such that objects which look similar should have a comparable redshift.

First of all, it is worth noting that this method performs quite well in
calculating the redshift for very different kinds of spectra. The only
requirement is that the density of reference objects is reasonably high in the
$d$-dimensional Euclidean space populated by the spectra. 
It could be shown that this method with its current set of reference
spectra (which is limited to redshifts $z \leq 0.5$, but the
reference sample is just densely populated until $z \approx 0.2$) can reach a
higher sensitivity than the SDSS pipeline for individual spectra. So far only the
completeness is considerably lower than in the SDSS pipeline, but this will be
improved via a larger and more representative reference sample which covers all redshifts.

To show the power of this new tool, in this work we presented outliers found
in the data set. For this a more conservative (more sensitive, but
less complete) parameter set has been chosen. We were able to detect outliers
by two different statistical redshifts:
The first approach focuses on the overall behavior of the spectra, thus being
less affected by confusion but being less informative.
The second approach is focusing on the behavior of predefined regions.
Its completeness rate is higher, i.e. more objects with exotic behavior have
been found. On the other hand the number of highlighted objects which appear
due to methodological artifacts is also increased. In summary both methods yield
very interesting objects where the SDSS redshift was wrong.

Even though these methods work quite nicely plenty of parameters exist which
are tunable and have impact on the final result. 
In the data pre-processing several models describing the continuum behavior
were investigated. The normalization of the spectra with respect to this
continuum and their noise might have an effect on the number of
true outliers, too. In addition the feature extraction has a severe impact on
the final results and might be tailored to certain scientific needs.



\subsection{Future Work}
In a next step we will investigate the impact of the 
choice of the reference sample. Each redshift bin should contain enough
reference objects to minimize systematical effects due to the bias of the
sample. This discussion is part of a forthcoming paper where the methodology is
applied to the full SDSS spectroscopic database.

In a final step the impact of the mathematical composition of the regression
values used in \mbox{Equation \ref{eq:median}} could be investigated. It would
further be interesting to study the behavior of different selection measure
such that a clearer distinction between noisy and good features can be made.
Additionally one could apply a pre- instead of a post-selection to
distinguish between signals and noise on the data level. This would 
make the reduction of the reference sample in the computational step 
easier, as just reference objects with an existing signal would be used for
comparison. On the other hand it would introduce further biases which have to
be tuned by the increased number of parameters. Some physical knowledge about
the type of signal which is expected would be required.

Finally the outlier detection could be modified. Depending on the scientific use
case the trade-off between completeness and sensitivity can be adjusted by
using different detection criteria.
Those detected outliers can be related to in future outlier
catalogs. As we are currently only investigating a small fraction of the
database ($<$1\%) a huge number of objects is expected to be marked as outliers
for the entire dataset, i.e. that the number of objects to be investigated will
be so large ($\approx$5,000) that a manual inspection will be extremely
time-consuming.
Anyway the discovery potential of this straightforward redshift determination
approach is huge. The applicability to compute model-independent redshifts of
new incoming data was already shown on this simplified and just partially
representative sub-sample.

%% file: Appendix.tex
\appendix
\thispagestyle{empty}
\section{Appendix}
\label{sec:app}
\begin{table}[ht!]
\centering
\input{finalTab.tex}
\end{table}

\newpage
\begin{figure}[ht!]
\centering\caption{The analysis of the H$_{\beta}$,$[$OIII$]$ region is shown.
One can see the relative difference in redshift against the SDSS redshift (top), the
distribution of the deviations (middle) and a histogram of the relative
difference in redshift (bottom). Noisy spectral features are marked in red, good
features in green.\label{pl1}}
\includegraphics[width=0.85\textwidth]{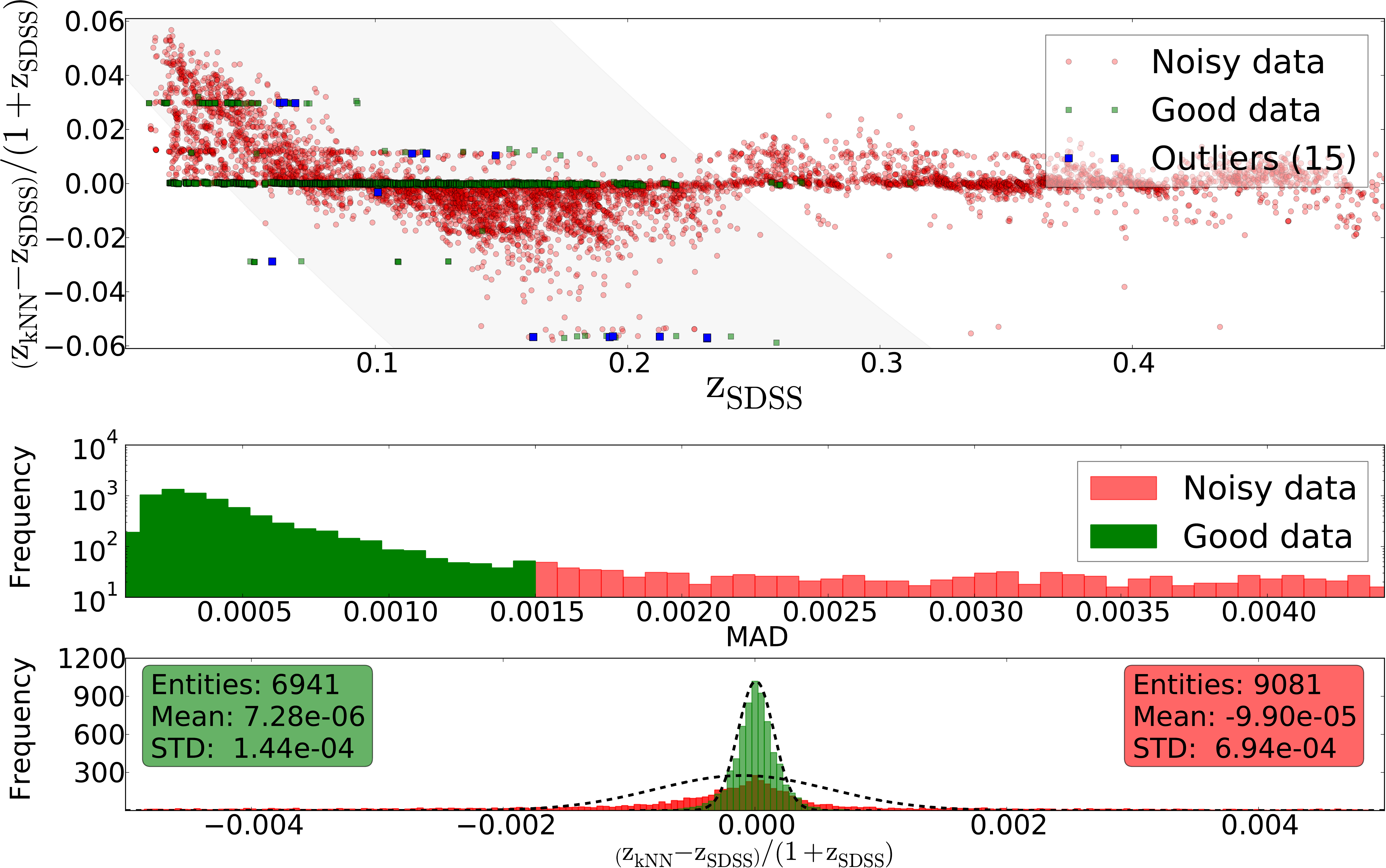}
\end{figure}
\begin{figure}[ht!]
\centering\caption{The analysis of the NaD region is shown. One can
see the relative difference in redshift against the SDSS redshift (top), the
distribution of the deviations (middle) and a histogram of the relative
difference in redshift (bottom). Noisy spectral features are marked in red, good
features in green.\label{pl2}}
\includegraphics[width=0.85\textwidth]{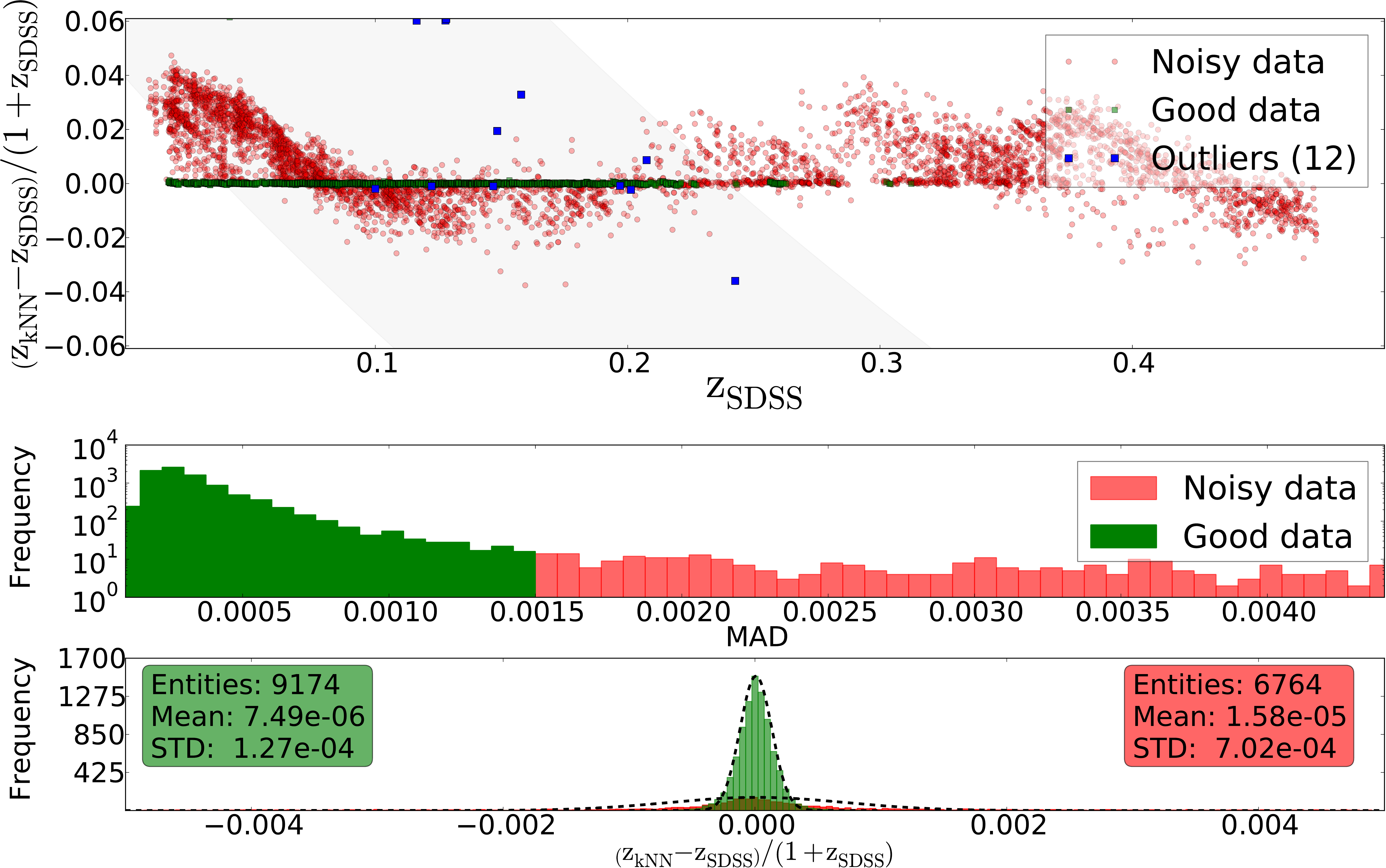}
\end{figure}

%% file: finalTab.tex
\caption{Summary of all manually investigated spectra. The
spectral features Em, Abs are from Experiment 1, all others
mark the respective regions where the feature was
detected as an outlier. Sorting is by number of spectral
features, spectral feature and finally identifier. \label{objSummary}}
\begin{tabular}{c|r|r|r|c|c|l}
Identifier & $z_{SDSS}$ & $z_{EST}$\hspace{2em} & $\Delta z$\hspace{0.75em} &
Spectral Feature & Class & Remarks \tabularnewline 
$[$plate-MJD-fiber$]$ & & & $[$km/s$]$ &  & & \tabularnewline
\hline
0266-51602-0095 & 0.0673 & 0.1266$\pm$0.0005 & 16600 & Em,H$_{\alpha}$,$[$NII$]$,$[$SII$]$ & true & Fig. \ref{fig:specialOnes} top \tabularnewline
0286-51999-0236 & 0.2075 & 0.0940$\pm$0.0007 & -28200 & Em,CaII,NaD & true &
possible lense, Fig. \ref{fig:specialOnes} center \tabularnewline 
0268-51633-0423 & 0.1699 & 0.1689$\pm$0.0005 & -300 & OII,H$_{\epsilon}$,H$_{\zeta}$ & true & shifted BL/NL \tabularnewline

0275-51910-0265 & 0.0596 & 0.0614$\pm$0.0013 & 500 & Em & wrong & low number density of \tabularnewline 
 & & & & & & reference objects (see Fig. \ref{fig:redDist})\tabularnewline
0279-51984-0449 & 0.2049 & 0.1264$\pm$0.0014 & -19600 & Em & wrong & member of G1\tabularnewline
0280-51612-0323 & 0.0576 & 0.0605$\pm$0.0007 & 800 & Em & wrong & low number density of \tabularnewline 
 & & & & & & reference objects (see Fig. \ref{fig:redDist})\tabularnewline
0282-51658-0493 & 0.2409 & 0.1600$\pm$0.0009 & -19600 & Em & wrong & member of
G1\tabularnewline

0267-51608-0601 & 0.0620 & 0.1709$\pm$0.0006 & 30700 & Abs & fake & fake
feature at $\lambda6,901$\tabularnewline 
0282-51630-0400 & 0.2690 & 0.1796$\pm$0.0008 & -21200 & Abs & true & Fig.
\ref{fig:specialOnes} bottom \tabularnewline

0274-51913-0617 & 0.0966 & 0.1159$\pm$0.0004 & 5200 & OII & true & dual core in image\tabularnewline
0272-51941-0332 & 0.2201 & 0.2184$\pm$0.0010 & -500 & H$_{\epsilon}$,H$_{\zeta}$
& true & NL shifted vs. absorption\tabularnewline 0288-52000-0215 & 0.1531 &
0.1543$\pm$0.0007 & 200 & H$_{\epsilon}$,H$_{\zeta}$ & true & shifted BL/NL\tabularnewline 
0268-51633-0354 & 0.0918 & 0.0928$\pm$0.0006 & 200 & H$_{\delta}$ & wrong & strong noise in spectral region \tabularnewline
0271-51883-0371 & 0.1202 & 0.0851$\pm$0.0012 & -9500 & H$_{\gamma}$ & wrong & litte active \& starburst spectra \tabularnewline
0273-51957-0579 & 0.1312 & 0.0946$\pm$0.0007 & -9800 & H$_{\gamma}$ & wrong & litte active \& starburst spectra \tabularnewline
0279-51608-0034 & 0.1010 & 0.0975$\pm$0.0014 & -1000 & H$_{\beta}$,$[$OIII$]$ &
wrong & high MAD \tabularnewline 
0271-51883-0570 & 0.0531 & 0.0638$\pm$0.0013 & 3000 & H$_{\alpha}$,$[$NII$]$ & wrong & low number density of \tabularnewline 
 & & & & & & reference objects (see Fig. \ref{fig:redDist})\tabularnewline
0286-51999-0089 & 0.1296 & 0.1263$\pm$0.0004 & -900 & H$_{\alpha}$,$[$NII$]$ &
wrong & very weak features only \tabularnewline 
0267-51608-0593 & 0.1631 & 0.1642$\pm$0.0005 & 200 & CaII & true & \tabularnewline
0275-51910-0142 & 0.1516 & 0.1526$\pm$0.0005 & 200 & CaII & true & \tabularnewline
0266-51602-0604 & 0.2995 & 0.3353$\pm$0.0012 & 8200 & Mgb & fake &
$\lambda6,913$ \tabularnewline 
0266-51630-0374 & 0.1661 & 0.1449$\pm$0.0007 & -5500 & Mgb & fake &
$\lambda5,892$\tabularnewline 
0270-51909-0537 & 0.1774 & 0.1589$\pm$0.0010 & -4800 & Mgb & wrong & QSO, sparse
in reference \tabularnewline 
0277-51908-0277 & 0.2822 & 0.2593$\pm$0.0013 & -5400 & Mgb & wrong & QSO, sparse
in reference \tabularnewline
0285-51930-0170 & 0.1794 & 0.1685$\pm$0.0009 & -2800 & Mgb & wrong & QSO, sparse
in reference \tabularnewline
0288-52000-0215 & 0.1531 & 0.1544$\pm$0.0006 & 300 & Mgb & true &
 shifted BL/NL \tabularnewline 
0266-51630-0318 & 0.1483 & 0.1706$\pm$0.0008 & 5800 & NaD & fake & 
$\lambda6,901$\tabularnewline 
0267-51608-0092 & 0.1467 & 0.1456$\pm$0.0005 & -300 & NaD & true & only NaD 
shifted \tabularnewline 
0267-51608-0320 & 0.1164 & 0.1836$\pm$0.0010 & 18000 & NaD & fake &
$\lambda6,976$ \tabularnewline 
0269-51910-0531 & 0.1278 & 0.1958$\pm$0.0011 & 18000 & NaD & fake &
$\lambda7,045$ \tabularnewline 
0270-51909-0114 & 0.1970 & 0.1960$\pm$0.0007 &
-300 & NaD & true & only NaD shifted \tabularnewline  
0274-51913-0548 & 0.1000 & 0.0979$\pm$0.0002 & -600 & NaD & true & only NaD
shifted \tabularnewline

0283-51660-0602 & 0.1281 & 0.1968$\pm$0.0007 & 18200 & NaD & fake &
$\lambda7,053$ \tabularnewline

0284-51943-0531 & 0.1578 & 0.1958$\pm$0.0005 & 9800 & NaD & fake &
$\lambda7,048$ \tabularnewline

0284-51943-0603 & 0.2013 & 0.1985$\pm$0.0006 & -700 & NaD & fake &
$\lambda7,062$ \tabularnewline

0285-51663-0602 & 0.2426 & 0.1979$\pm$0.0007 & -10800 & NaD & fake &
$\lambda7,058$ \tabularnewline 
0285-51930-0035 & 0.1222 & 0.1212$\pm$0.0002 &
-300 & NaD & true & only NaD shifted \tabularnewline
\end{tabular}